\documentclass[12pt]{amsart}
\usepackage{amsmath}
\usepackage{amsthm}
\usepackage{amsfonts}
\usepackage{graphicx}
\usepackage{amssymb} %for \geqslant
\newenvironment{nouppercase}{
  
  \renewcommand{\uppercasenonmath}[1]{}}{}

\begin{document}

\title[Integrability in the dynamics of axially symmetric membranes]
{Integrability in the dynamics of axially symmetric membranes}
\author[Jens Hoppe]{Jens Hoppe}
%\date{17 Dez 2021}
\address{Braunschweig University, Germany}
\email{jens.r.hoppe@gmail.com}

\begin{abstract}
B\"acklund-type transformations in four-dimensional space-time and an intriguing reduced zero-curvature formulation for axially symmetric membranes, with diffeomorphism- resp. Lorentz- symmetries reappearing after orthonormal gauge-fixing, are found.  
\end{abstract}

\begin{nouppercase}
\maketitle
\end{nouppercase}
\thispagestyle{empty}
%%%%%%%%%%%%%%%%%%%%%%%%%%%%%%%%%%%%%%%%%%%%%%%%%%%%%%%%%%%%%%%%%%%%%%%%%%%%%
%PAGE 1 PAGE 1 PAGE 1 PAGE 1 PAGE 1 PAGE 1 PAGE 1 PAGE 1 PAGE 1 PAGE 1 PAGE 1 
%%%%%%%%%%%%%%%%%%%%%%%%%%%%%%%%%%%%%%%%%%%%%%%%%%%%%%%%%%%%%%%%%%%%%%%%%%%%%
\noindent
Axially symmetric membranes in $\mathbb{R}^{1,3}$ are the simplest case in a notoriously difficult class of problems that have long withstood to be solved. In \cite{H82}, an orthonormally gauge fixed light-cone formulation was given (in terms of only one field, with polynomial Hamiltonian and no explicit gange-fixing constraint remaining), in \cite{BH93} an important relation with hydrodynamics was found
% (see also \cite{H94},\cite{Y94}), 
and in \cite{H94} the first non-trivial explicit solutions (after the `obvious' spherically symmetric one, 40 years ago \cite{D62}). The existence of transformations similar to those of B\"acklund and Bianchi \cite{BB1899}, resp. Thybaut \cite{T1897}, was conjectured in \cite{H14} \cite{Y94}, 
%%%%%%%%%%%%%%%%%%%%%%%%%%%%%%%%%%%%%%%%%%%%%%%%%%%%%%%%%%%%%%%%%%%%%%%%%%%%%
%PAGE 1' PAGE 1' PAGE 1' PAGE 1' PAGE 1' PAGE 1' PAGE 1' PAGE 1' PAGE 1' PAGE 1'
%%%%%%%%%%%%%%%%%%%%%%%%%%%%%%%%%%%%%%%%%%%%%%%%%%%%%%%%%%%%%%%%%%%%%%%%%%%%%
and several non-trivial observations concerning axially symmetric membranes were mentioned in part IV of \cite{H19}. Recently \cite{H22}, in the context of new explicit (infinite-energy) solutions \cite{H21}, an astonishingly simple transformation between different orthonormal parametrizations was observed. These transformations (discussed thoroughly in the first part of the present paper), mapping solutions (of different PDE's) into each other (at first describing the {\it same} zero mean curvature world-volume $\mathfrak{M}_3 \subset \mathbb{R}^{1,3}$) are similar to the above mentioned B\"acklund transformations, that originally were used in the context of generating 2-dimensional constant negative curvature surfaces (`in $\mathbb{R}^3$').\\
%%%%%%%%%%%%%%%%%%%%%%%%%%%%%%%%%%%%%%%%%%%%%%%%%%%%%%%%%%%%%%%%%%%%%%%%%%%%%
%PAGE 2 PAGE 2 PAGE 2 PAGE 2 PAGE 2 PAGE 2 PAGE 2 PAGE 2 PAGE 2 PAGE 2 PAGE 2 
%%%%%%%%%%%%%%%%%%%%%%%%%%%%%%%%%%%%%%%%%%%%%%%%%%%%%%%%%%%%%%%%%%%%%%%%%%%%%
One way (see e.g.\cite{H14}) to describe axially symmetric membrane motions in 4-dimensional Minkowski-space is via solutions ($r(t,\varphi)$, $z(t,\varphi)$) of 
\begin{equation}\label{eq1} 
\dot{r}r' + \dot{z}z' = 0, \qquad \dot{r}^2 + \dot{z}^2 + \frac{r^2(r'^2 + z'^2)}{\rho^2} = 1
\end{equation}
(where $\boldsymbol{\dot{•}}$ and $'$ denote differentiation with respect to time $t$ and `angle' $\varphi$), $\rho(\varphi)$ may be chosen to be constant, $\rho = \rho_0 = r^2_0$, (by reparametrizing $\varphi \rightarrow \tilde{\varphi}(\varphi)$, $\frac{d\tilde{\varphi}}{d\varphi} = \rho(\varphi)$; in some cases, cp. \cite{H22},paying the price of having non-compact ranges), which (by scaling $t$, $r$ and $z$) could then allow one to put $\rho = 1$; hence describing membrane solutions, resp. the 3-dimensional world-volume swept out in space-time, by
\begin{equation}\label{eq2} 
\begin{array}{l}
x^{\mu}(t,\varphi,\psi) = 
\begin{pmatrix}
t\\ r(t,\varphi)\cos \psi \\ r(t,\varphi)\sin \psi \\ z(t,\varphi)
\end{pmatrix} = 
\begin{pmatrix}
t\\ \vec{x}(t,\varphi)
\end{pmatrix}\\[1cm]
(\dot{r} \pm \frac{rr'}{r^2_0})^2 + (\dot{z} \pm \frac{rz'}{r^2_0})^2 = 1
\end{array}
\end{equation}
with $r_0 = 1$.\\[0.15cm]
As long as ${\dot{r}\choose \dot{z}}$  and ${r' \choose z'}$,
%%%%%%%%%%%%%%%%%%%%%%%%%%%%%%%%%%%%%%%%%%%%%%%%%%%%%%%%%%%%%%%%%%%%%%%%%%%%%
%PAGE 2' PAGE 2' PAGE 2' PAGE 2' PAGE 2' PAGE 2' PAGE 2' PAGE 2' PAGE 2' PAGE 2' 
%%%%%%%%%%%%%%%%%%%%%%%%%%%%%%%%%%%%%%%%%%%%%%%%%%%%%%%%%%%%%%%%%%%%%%%%%%%%%
velocity- and tangent-vector of a time dependent curve in $\mathbb{R}^2$, are linearly independent, (\ref{eq1}) implies the second-order `membrane' equations
\begin{equation}\label{eq3} 
r^4_0\ddot{z} = (r^2z')', \qquad r^4_0\ddot{r} = (r^2r')' - r(r'^2+z'^2),
\end{equation}
consistent with
\begin{equation}\label{eq4} 
\begin{array}{l}
\big(G_{\alpha\beta}\big)  = 
\begin{pmatrix}
1-\dot{r}^2-\dot{z^2} & 0 & 0\\
0 & -(r'^2+z'^2) & 0 \\
0 & 0 & -r^2
\end{pmatrix} =
\bigg(\dfrac{\partial x^{\mu}}{\partial \varphi^{\alpha}} \dfrac{\partial x^{\nu}}{\partial \varphi^{\beta}}\eta_{\mu\nu} \bigg)_{\alpha\beta = 0,1,2}\\[1cm]
\Delta x^{\mu} = \frac{1}{\sqrt{G}} \partial_{\alpha}\sqrt{G}G^{\alpha\beta}\partial_{\beta}x^{\mu}   =
\frac{1}{\sqrt{G}}\big( \partial^2_t - \partial_{\varphi}\frac{r^2}{r^4_0}  \partial_{\varphi} - (\frac{r'^2 + z'^2}{r^4_0} )\partial^2_{\psi} \big)x^{\mu}
 = 0.  
\end{array}
\end{equation}
The (at first astonishing) fact that the second-order equations (\ref{eq3}) follow from the first-order `gauge-conditions' (\ref{eq1}) has to do with $\Delta x^{\mu}$ always being purely normal to the world-volume $\mathfrak{M}_3$ ($\partial_{\alpha}x_{\mu}\cdot \Delta x^{\mu} \equiv 0$),i.e. the 4 equations $\Delta x^{\mu} = 0$ having only one independent  component - which can be taken to be $\frac{1}{3}n_{\mu} \Delta x^{\mu} = \frac{1}{3}G^{\alpha\beta} \cdot \big(H_{\alpha\beta} := n_{\mu}\partial^2_{\alpha\beta} x^{\mu}\big)$, the mean-curvature of $\mathfrak{M}_3$ ,
%%%%%%%%%%%%%%%%%%%%%%%%%%%%%%%%%%%%%%%%%%%%%%%%%%%%%%%%%%%%%%%%%%%%%%%%%%%%%
%PAGE 2'' PAGE 2'' PAGE 2'' PAGE 2'' PAGE 2'' PAGE 2'' PAGE 2'' PAGE 2'' PAGE 2''  
%%%%%%%%%%%%%%%%%%%%%%%%%%%%%%%%%%%%%%%%%%%%%%%%%%%%%%%%%%%%%%%%%%%%%%%%%%%%%
but (generically equivalently) also (only/simply) as {\it one} of the 4 equations; due to the choice of $\varphi^0$ and the orthogonal parametrization provided by the first part of (\ref{eq1}) (as well as the $U(1)$-symmetry) $\Delta x^0 = 0$ becomes a (first order) conservation law, $\partial_t(\sqrt{G}G^{00}) = 0$, which is the second part of (\ref{eq1})-the local conserved quantity being $\rho$ (which is the energy-density $\mathcal{H}$ in a Hamiltonian O(rtho)N(ormal) `3+1' description of membranes \cite{12} \cite{M06} \cite{ZT10} \cite{H13}).\\
While `integrability' of the above equations has been suspected for quite some time (cp.\cite{Y94},\cite{H14},\cite{EHHS15}, \cite{H19}), the present paper reveals more concrete, {\it substantial}, signs of integrability.
%%%%%%%%%%%%%%%%%%%%%%%%%%%%%%%%%%%%%%%%%%%%%%%%%%%%%%%%%%%%%%%%%%%%%%%%%%%%%
%PAGE 3 PAGE 3 PAGE 3 PAGE 3 PAGE 3 PAGE 3 PAGE 3 PAGE 3 PAGE 3 PAGE 3 PAGE 3 
%%%%%%%%%%%%%%%%%%%%%%%%%%%%%%%%%%%%%%%%%%%%%%%%%%%%%%%%%%%%%%%%%%%%%%%%%%%%%  
\section{(B\"acklund-) Transformations between Different Orthonormal Parametrizations}
\noindent
Instead of choosing $\varphi^0 = x^0$ (= the time $t$ of a Lorentz-observer), choose $\varphi^0$ to be $x^3$, i.e. 
\begin{equation}\label{eq5}
\begin{array}{l} 
\tilde{x}^{\mu}(u,v,w) = 
\begin{pmatrix}
T(u,v)\\ R(u,v) {\cos w \atop \sin w}\\ u
\end{pmatrix} = 
\begin{pmatrix}
\tilde{x}^{\alpha}(u,v,w) \\ u
\end{pmatrix},\\[0.75cm]
\tilde{G}_{\alpha\beta} = \frac{\partial \tilde{x}^{\mu}}{\partial \varphi^{\alpha}}\frac{\partial \tilde{x}^{\nu}}{\partial \varphi^{\beta}}\eta_{\mu\nu}
\end{array}
\end{equation}
and again an orthogonal parametrization (i.e. $v = \varphi^1$ such that $\tilde{G}_{01} = 0$; $\tilde{G}_{02} = 0 = \tilde{G}_{12}$ because of the $u(1)$-symmetry; implying $\partial_u(\tilde{\rho}: = \sqrt{\tilde{G}}\tilde{G}^{00})=0$ if (\ref{eq5}) describes a stationary point of the volume-functional).\\
In \cite{H22} the particular (membrane) solution (stationary point)
\begin{equation}\label{eq6} 
\begin{array}{l}
T(u,v)  = H(u)\cdot v, \quad
R(u,v)  = H(u)\sqrt{v^2 - 1} \\[0.25cm]
H'^2  = H^4 +1, \quad \tilde{x}^{\alpha}  = H(u) {v \choose \sqrt{v^2 - 1}{ \cos w \atop \sin w}}\\
\end{array}
\end{equation}
(in order to have $\tilde{\rho} = 1 = \rho$, $v$ here equals $\cosh v$ in \cite{H22}, and $\varphi$ here corresponds to $\frac{1}{2}\varphi^2$, there) was found; note that \eqref{eq6} is {\it not} a reparametrization of Dirac's $SO(3)$ invariant solution, $\vec{x}(t, \varphi, \psi) = D(t){\sqrt{1-\varphi^2}{ \cos \psi \atop \sin \psi} \choose \varphi}_{\varphi \in [-1,+1]}$, $ \dot{D}^2 + D^4 = 1$, but its $SO(2,1)$ invariant `cousin'.
%%%%%%%%%%%%%%%%%%%%%%%%%%%%%%%%%%%%%%%%%%%%%%%%%%%%%%%%%%%%%%%%%%%%%%%%%%%%%
%PAGE 3' PAGE 3' PAGE 3' PAGE 3' PAGE 3' PAGE 3' PAGE 3' PAGE 3' PAGE 3' PAGE 3' 
%%%%%%%%%%%%%%%%%%%%%%%%%%%%%%%%%%%%%%%%%%%%%%%%%%%%%%%%%%%%%%%%%%%%%%%%%%%%%  
The general orthonormality conditions and corresponding equations of motion for \eqref{eq5}, in complete analogy to \eqref{eq1} and \eqref{eq3}, are
\begin{equation}\label{eq7} 
T_u T_v - R_u R_v = 0, \quad T^2_u-R^2_u -1 = \frac{R^2(R^2_v -T^2_v)}{R^4_0}
\end{equation}
\begin{equation}\label{eq8} 
R^4_0T_{uu} = (R^2T_v)_v, \quad R^4_0R_{uu} = \underset{(= R^2R_{vv} + RR^2_v + R T^2_v)}{(R^2R_v)_v - R(R^2_v - T^2_v)}.
\end{equation}
While it is obvious that \eqref{eq5} and \eqref{eq2}, for one and the same world-volume, must give $R(u,v) = r(t,\varphi)$, $t = T(u,v)$, $u=z(t,\varphi)$, with some concrete relation $\varphi = \phi(u,v)$ resp. $v = v(t,\varphi)$, the important observation, in the example (\ref{eq6}) e.g. allowing to conclude
\begin{equation}\label{eq9} 
\phi(u,v) = \sqrt{H^4+1}\frac{1}{2}(v^2-1) =: \varphi
\end{equation}
(corresponding to $\frac{1}{2}\varphi^2$ in \cite{H22}) is that the (`B\"acklund') transformations between the two triply orthogonal parametrizations of $\mathfrak{M}_3$, are given by first order {\it linear} PDE's, following from
%%%%%%%%%%%%%%%%%%%%%%%%%%%%%%%%%%%%%%%%%%%%%%%%%%%%%%%%%%%%%%%%%%%%%%%%%%%%%
%PAGE 3'' PAGE 3'' PAGE 3'' PAGE 3'' PAGE 3'' PAGE 3'' PAGE 3'' PAGE 3'' PAGE 3'' 
%%%%%%%%%%%%%%%%%%%%%%%%%%%%%%%%%%%%%%%%%%%%%%%%%%%%%%%%%%%%%%%%%%%%%%%%%%%%% 
\begin{equation}\label{eq10} 
\begin{split}
\tilde{G}_{\alpha \beta} & = 
\begin{pmatrix}
T^2_u-R^2_u -1 & 0 & 0 \\ 
0 & -(R^2_v-T^2_v) & 0 \\
0 & 0 & -R^2
\end{pmatrix}\\[0.25cm]
& = J^T
\begin{pmatrix}
1-\dot{r}^2-\dot{z}^2 & 0 & 0 \\ 
0 & -(r'^2 + z'^2) & 0 \\
0 & 0 & -r^2
\end{pmatrix} J \\[0.25cm]
J & = \frac{\partial t, \varphi, \psi}{\partial u \; v \; w} =
\begin{pmatrix}
T_u & T_v & 0 \\ 
\phi_u & \phi_v & 0 \\
0 & 0 & 1
\end{pmatrix},
\end{split}
\end{equation}
implying in particular $G_{00}T_uT_v + G_{11}\phi_u\phi_v = 0$, $\tilde{G}_{00}\dot{u}u' + \tilde{G}_{11}\dot{v}v' = 0$.
%%%%%%%%%%%%%%%%%%%%%%%%%%%%%%%%%%%%%%%%%%%%%%%%%%%%%%%%%%%%%%%%%%%%%%%%%%%%%
%PAGE 4 PAGE 4 PAGE 4 PAGE 4 PAGE 4 PAGE 4 PAGE 4 PAGE 4 PAGE 4 PAGE 4 PAGE 4 
%%%%%%%%%%%%%%%%%%%%%%%%%%%%%%%%%%%%%%%%%%%%%%%%%%%%%%%%%%%%%%%%%%%%%%%%%%%%% 
Due to
\begin{equation}\label{eq11} 
\begin{split}
1 & = \frac{\rho^2}{r^4_0} = \frac{\frac{r^2}{r^4_0}(r'^2+z'^2)}{1-\dot{r}^2-\dot{z^2}} = \frac{G_{22}G_{11}}{r^4_0G_{00}} 
= 1 = \frac{\tilde{G}_{22}\tilde{G}_{11}}{R^4_0\tilde{G}_{00}}\\[0.15cm]
& = \frac{\frac{R^2}{R^4_0}(R^2_v-T^2_v)}{T^2_u-R^2_u-1} 
= \frac{\tilde{\rho}^2}{R^4_0} = 1
\end{split}
\end{equation}
(hence $r^4_0\frac{G_{00}}{G_{11}} = G_{22} = -r^2(t,\varphi) = -R^2(u,v) = \tilde{G}_{22} = \frac{\tilde{G}_{00}R^4_0}{\tilde{G_{11}}} $), \eqref{eq10} implies
\begin{equation}\label{eq12} 
\begin{split}
r^4_0\phi_u \phi_v & = r^2T_uT_v = R^2R_uR_v\\
R^4_0 \dot{v}v' & = R^2\dot{u}u' = r^2\dot{z}z' = -r^2\dot{r}r'
\end{split}
\end{equation}
i.e.
\begin{equation}\label{eq13} 
R^4_0\frac{T_u\phi_u}{T_v\phi_v} = R^4_0\frac{\dot{v}v'}{\dot{u}u'} = R^2 = r^2 = r^4_0\frac{\phi_u\phi_v}{T_uT_v} = \frac{\dot{u}\dot{v}}{u'v'}r^4_0,
\end{equation}
hence
\begin{equation}\label{eq14} 
\dot{u}^2 = v'^2\frac{R^4_0}{r^4_0}, \quad T^2_u = \phi^2_v\cdot\frac{r^4_0}{R^4_0}
\end{equation}
when using that, given the transformation 
\begin{equation}\label{eq15} 
\begin{array}{c}
\big(z(t,\varphi) = u, v\big) \leftrightarrow \big(T(u,v) = t, \; \phi(u,v) = \varphi\big)\\[0.25cm]
\begin{pmatrix}
T_u & T_v \\ \phi_u & \phi_v
\end{pmatrix} =
\begin{pmatrix}
\dot{u} & u' \\ \dot{v} & v'
\end{pmatrix}^{-1} =
\dfrac{1}{(\delta := \dot{u}v' -u'\dot{v})}
\begin{pmatrix}
v' & -u' \\ -\dot{v} & \dot{u}
\end{pmatrix},
\end{array}
\end{equation}
simplifying \eqref{eq12} even further (while making some sign choice/s for $\phi$ and $v$),
\begin{equation}\label{eq16} 
\begin{array}{l}
\frac{r^2_0}{R^2_0}\phi_v  = T_u, \; r^2_0\phi_u = \frac{R^2}{R^2_0}T_v; \\[0.15cm]
\dot{u}  = v'\frac{R^2_0}{r^2_0}, \; R^2_0\dot{v} = \frac{r^2}{r^2_0}u'
\end{array}
\end{equation}
(which itself implies $T_{uu} = (\frac{R^2}{R^4_0}T_v)_v$, resp. $\ddot{z} = (\frac{r^2}{r^4_0}z')'$; note that every appearance of $\varphi$ resp. $\phi$ goes with a factor $r^2_0$, every appearance of $v$ with a factor $R^2_0$).
%%%%%%%%%%%%%%%%%%%%%%%%%%%%%%%%%%%%%%%%%%%%%%%%%%%%%%%%%%%%%%%%%%%%%%%%%%%%%
%PAGE 5 PAGE 5 PAGE 5 PAGE 5 PAGE 5 PAGE 5 PAGE 5 PAGE 5 PAGE 5 PAGE 5 PAGE 5 
%%%%%%%%%%%%%%%%%%%%%%%%%%%%%%%%%%%%%%%%%%%%%%%%%%%%%%%%%%%%%%%%%%%%%%%%%%%%% 
Solving the linear first order PDE's for $\phi(u,v)$ (resp. $v(t,\varphi)$) then provides a (`B\"acklund') map between solutions ($T(u,v), R(u,v)$) satisfying \eqref{eq7}/\eqref{eq8} and ($r(t,\varphi), z(t,\varphi)$) satisfying \eqref{eq1}/\eqref{eq3} - which one can easily prove as follows: suppose e.g. $T$ and $R$ satisfy \eqref{eq7}, and $\phi$ is a solution of \eqref{eq16}; define $r(t,\varphi) := R(u(t,\varphi), v(t,\varphi))$, where $(u,v)\leftrightarrow (t = T(u,v), \varphi = \phi(u,v))$. Then one first straightforwardly verifies that 
\begin{equation}\label{eq17} 
\begin{split}
\dot{r}r' + \dot{z}z' & = (\dot{u}R_u + \dot{v}R_v)(u'R_u + v'R_v) + \dot{u}u' \\
& = \delta \big( (\phi_vR_u - \phi_uR_v)(-T_vR_u + T_uR_v) - T_v\phi_v  \big) \\
& = \frac{R^2_0}{r^2_0}\delta \big( (T_uR_u - \frac{R^2}{R^4_0} R_vT_v)(-T_vR_u + T_uR_v) - T_uT_v  \big) = \ldots = 0.
\end{split}
\end{equation}
\eqref{eq10} on the other hand implies
\begin{equation}\label{eq18} 
\begin{split}
\frac{r'^2 + z'^2}{r_0^4} & = \frac{R^2_v - T^2_v}{R^4_0T^2_u - R^2 T^2_v} \big(= \frac{1}{r^2_0R^2_0} \, \frac{R^2_v-T^2_v}{|J|}\big)\\[0.15cm]
1-\dot{r}^2 - \dot{z}^2 & =  \frac{T^2_u - R^2_u - 1}{T^2_u - \frac{R^2}{R^4_0} T^2_v} \big( = \frac{R^2_0}{r^2_0} \, \frac{T^2_u-R^2_u-1}{|J|}\big).
\end{split}
\end{equation}
%%%%%%%%%%%%%%%%%%%%%%%%%%%%%%%%%%%%%%%%%%%%%%%%%%%%%%%%%%%%%%%%%%%%%%%%%%%%%
%PAGE 6 PAGE 6 PAGE 6 PAGE 6 PAGE 6 PAGE 6 PAGE 6 PAGE 6 PAGE 6 PAGE 6 PAGE 6 
%%%%%%%%%%%%%%%%%%%%%%%%%%%%%%%%%%%%%%%%%%%%%%%%%%%%%%%%%%%%%%%%%%%%%%%%%%%%% 
While the above relations/transformations between the two orthonormal parametrizations are very nice, they do not a priori help to find new solutions. One should look for symmetries that one could use (on either of the two sides) which can help to generate from one (original/simple) solution `most' of the other solutions by a mixture of solving the linear first-order PDE(s) \eqref{eq16}, and applying symmetry transformations. As it is rather tedious to carry along the factors of $r_0$ and $R_0$, both will now be put $=1$ (until specifically needed) as was already done\footnote{It is important to note that there are two very different kinds of transformations, which effect $r_0 (R_0)$, hence could be used to make them $=1$: scaling of all space-time coordinate, $x^{\mu}\rightarrow \lambda x^{\mu}$ (resulting in $r_0 \rightarrow \lambda r_0$ resp. $R_0 \rightarrow \lambda R_0$; explaining why the conserved quantity $\rho_0 = \int \rho (\varphi) d\varphi$ was chosen to be denoted by the {\it square} of the constants $r_0$ and $R_0$)-- for any given solution, such a scaling (changing the actual/physical/geometric solution) may be used to put the conserved quantity $=1$; or: changing the scale of $v$ or $\varphi$ (which both are reparametrizations, i.e. {\it not} changing the geometric/physical solution; hence $\rho = 1 = \tilde{\rho}$ is no real loss of generality, but for compact membranes one, in some cases, has to be careful to not generate multicovering)}
in the explicit $(u,v)$ example \eqref{eq6} (cp. \cite{H21}/\cite{H22}), $T(u,v) = H(u)\cdot v$, $R(u,v) = H(u)\sqrt{v^2 -1}$, where $\phi(u,v) = \sqrt{H^4+1}\frac{1}{2}(v^2-1)$, cp.\eqref{eq9}, resp. $u=(t,\varphi)$ and $v$ related to $t$ and $\varphi = \phi(u,v)$ by 
\begin{equation}\label{eq19} 
\frac{2\varphi}{\sqrt{1+H^4}} = \frac{t^2}{H^2}-1, \quad \frac{4\varphi^2}{(v^2-1)^2} - \frac{t^4}{v^4} = 1,
\end{equation}
and $r(t,\varphi) = R(u(t,\varphi), v(t,\varphi))$.
%%%%%%%%%%%%%%%%%%%%%%%%%%%%%%%%%%%%%%%%%%%%%%%%%%%%%%%%%%%%%%%%%%%%%%%%%%%%%
%PAGE 7 PAGE 7 PAGE 7 PAGE 7 PAGE 7 PAGE 7 PAGE 7 PAGE 7 PAGE 7 PAGE 7 PAGE 7 
%%%%%%%%%%%%%%%%%%%%%%%%%%%%%%%%%%%%%%%%%%%%%%%%%%%%%%%%%%%%%%%%%%%%%%%%%%%%% 
\section{Reduced, Single-field, Parametric description}
\noindent
As the second order equations for $T(u,v)$ resp. $z(t,\varphi)$ arise as compatibility/integrability equations of first-order equations, one may wonder whether this is also true for $r(t,\varphi) = R(u,v)$.
As \eqref{eq7} implies 
\begin{equation}\label{eq20} 
\begin{array}{l}
(\phi_v =) T_u  = F(R, R_u, R_v) = F[R](= Y_{uv}) \\[0.15cm]
(\frac{1}{R^2} \phi_u = Y_{vv} =) T_v  = \frac{R_uR_v}{F}
\end{array}
\end{equation}
with
\begin{equation}\label{eq21} 
\begin{split}
F^2_{\pm} & = \frac{1}{2}(R^2_u + R^2R^2_v+1) \pm \sqrt{\frac{1}{4}(R^2_u + R^2R^2_v+1)^2 - R^2R^2_uR^2_v} \\
 & =: P \pm \sqrt{P^2 - Q^2} \geqslant 0 
\end{split}
\end{equation}
cross-differentiation, 
\begin{equation}\label{eq22} 
(F)_v = \bigg( \frac{R_u R_v}{F}\bigg)_u, \quad (F)_u = \bigg( \frac{R^2R_u R_v}{F}\bigg)_v
\end{equation}
{\it will} result in  a second-order PDE for $R$. The consequences of \eqref{eq21} (allowing for four different choices of $F$) require a careful discussion, which will now follow for the corresponding equations,
\begin{equation}\label{eq23} 
\begin{split}
\dot{z} & = f[r]\\
z' & = \frac{-\dot{r}r'}{f[r]}
\end{split}
\end{equation}
%%%%%%%%%%%%%%%%%%%%%%%%%%%%%%%%%%%%%%%%%%%%%%%%%%%%%%%%%%%%%%%%%%%%%%%%%%%%%
%PAGE 8 PAGE 8 PAGE 8 PAGE 8 PAGE 8 PAGE 8 PAGE 8 PAGE 8 PAGE 8 PAGE 8 PAGE 8  
%%%%%%%%%%%%%%%%%%%%%%%%%%%%%%%%%%%%%%%%%%%%%%%%%%%%%%%%%%%%%%%%%%%%%%%%%%%%% 
 \begin{equation}\label{eq24} 
\begin{split}
f^2_{\pm} & = \frac{1}{2}(1-\dot{r}^2-r^2r'^2) \pm \sqrt{\frac{1}{2}(1-\dot{r}^2-r^2r'^2)^2 - r^2\dot{r}^2r'^2}\\
& =: p \pm \sqrt{p^2 - q^2}.
\end{split}
\end{equation}
A straightforward, but astonishingly lengthy/tedious calculation of $2f(f' + (\frac{r'\dot{r}}{f})\dot{•}\,)$ 
gives $\frac{\pm r'}{\sqrt{p^2 - q^2}}$ times the LHS of 
\begin{equation}\label{eq25} 
(\ddot{r}-r^2r''-rr'^2)(p\pm \sqrt{p^2-q^2}) + \frac{\dot{r}^2}{r}(p\mp \sqrt{p^2-q^2}) + \dot{r}^2(\ddot{r}-r^2r'') = 0.
\end{equation}
Denoting $(\ddot{r}-r^2r''-rr'^2)$ by $D_0[r]$, and using 
\begin{equation}\label{eq26} 
f_{\pm}f_{\mp} = r\dot{r}r' = q
\end{equation}
\eqref{eq25} can be written as
\begin{equation}\label{eq27} 
(\dot{r}^2 + (p\pm \sqrt{p^2-q^2}))(D_0[r] + \frac{1}{r}(p\mp \sqrt{p^2-q^2})) = 0,
\end{equation}
which implies
\begin{equation}\label{eq28} 
D_{\pm}[r] := \ddot{r} - r^2r'' - rr'^2 + \frac{1}{r}(p \mp \sqrt{p^2 - q^2}) = 0,
\end{equation}
which {\it is} the correct second-order PDE for $r(t,\varphi)$, as the last term equals $rz'^2$.
%%%%%%%%%%%%%%%%%%%%%%%%%%%%%%%%%%%%%%%%%%%%%%%%%%%%%%%%%%%%%%%%%%%%%%%%%%%%%
%PAGE 8' PAGE 8' PAGE 8' PAGE 8' PAGE 8' PAGE 8' PAGE 8' PAGE 8' PAGE 8' PAGE 8' 
%%%%%%%%%%%%%%%%%%%%%%%%%%%%%%%%%%%%%%%%%%%%%%%%%%%%%%%%%%%%%%%%%%%%%%%%%%%%% 
An analogous (similarly tedious, equally intricate) calculation, $f^2 = p \pm \sqrt{p^2 - q^2}$, gives $ \frac{2f}{\dot{r}}\sqrt{p^2-q^2}(\dot{f}+ (r\tilde{f})')$ $= \underset{(+)}{-} (D_0[r] + \frac{1}{r}(p\mp \sqrt{p^2 - q^2}))(r^2r'^2 + (p\pm \sqrt{p^2-q^2}))$.\\
%%%%%%%%%%%%%%%%%%%%%%%%%%%%%%%%%%%%%%%%%%%%%%%%%%%%%%%%%%%%%%%%%%%%%%%%%%%%%
%PAGE 9 PAGE 9 PAGE 9 PAGE 9 PAGE 9 PAGE 9 PAGE 9 PAGE 9 PAGE 9 PAGE 9 PAGE 9 
%%%%%%%%%%%%%%%%%%%%%%%%%%%%%%%%%%%%%%%%%%%%%%%%%%%%%%%%%%%%%%%%%%%%%%%%%%%%% 
For definiteness take
\begin{equation}\label{eq29} 
f_{\pm} = \frac{1}{2}\bigg( \sqrt{1-(\dot{r}-rr')^2} \pm \sqrt{1-(\dot{r}+rr')^2} \bigg) =: \frac{1}{2}(\sqrt{-} \pm \sqrt{+});
\end{equation}
multiplying \eqref{eq29} by an overall minus-sign would not change \eqref{eq26}, nor \eqref{eq24}, but ($\frac{-\dot{r}r'}{f_{\pm}} = \frac{-f_{\mp}}{r}$, cp.\eqref{eq26}) simply correspond to $z\rightarrow -z$, consistent with
\begin{equation}\label{eq30} 
\begin{split}
\dot{z} + rz' & = f_{\pm} - f_{\mp} = \pm \sqrt{1-(\dot{r}+rr')^2}\\
\dot{z} - rz' & = f_{\pm} + f_{\mp} =  \sqrt{1-(\dot{r}-rr')^2}.
\end{split}
\end{equation}
Note that if $r = r_+$ is a solution of $D_+[r_+] = 0$ (with $z = z_+$ then obtained from $\dot{z}_+ = f_+[r_+]$, $z'_+ = -\frac{1}{r}f_-[r_+]$) $r_+$ will generically {\it not} satisfy $D_-[r_+] = 0$; so the somewhat curious situation occurs that the solutions of \eqref{eq23} resp. (cp.\eqref{eq3}) $D_z[r] := \ddot{r} - r^2r'' - rr'^2 + rz'^2 = 0$ fall into two (more or less disjoint) sets, ($r_+, z_+$) and ($r_-, z_-$).
%%%%%%%%%%%%%%%%%%%%%%%%%%%%%%%%%%%%%%%%%%%%%%%%%%%%%%%%%%%%%%%%%%%%%%%%%%%%%
%PAGE 10 PAGE 10 PAGE 10 PAGE 10 PAGE 10 PAGE 10 PAGE 10 PAGE 10 PAGE 10 PAGE 10  
%%%%%%%%%%%%%%%%%%%%%%%%%%%%%%%%%%%%%%%%%%%%%%%%%%%%%%%%%%%%%%%%%%%%%%%%%%%%% 
\section{Characteristic Coordinates}
\noindent
In \cite{H19} many different aspects of `characteristic coordinates' were discussed. Here, it is perhaps simplest to say that one can e.g. use the differmorphism invariance of the Volume-functional to choose coordinates $\theta_+, \theta_-, \theta$ on the $U(1)$ invariant world-volume $\mathfrak{M}_3$ to choose the $2$ tangent vectors $x_{\pm} := \frac{\partial x}{\partial \theta_{\pm}}$ to be null, i.e.
\begin{equation}\label{eq31} 
t^2_{\pm} - r^2_{\pm} - z^2_{\pm} = 0.
\end{equation}
Then any of the 3 conditions
\begin{equation}\label{eq32} 
\partial_+(rt_-) + \partial_-(rt_+) = 0,
\end{equation}
\begin{equation}\label{eq33} 
\partial_+(rz_-) + \partial_-(rz_+) = 0,
\end{equation}
\begin{equation}\label{eq34} 
\partial_+(rr_-) + \partial_-(rr_+) + x_+x_- = 0,
\end{equation}
will guarantee that $\mathfrak{M}_3$ has zero mean curvature. The metric tensor, its determinant, and its inverse in these coordinates are 
\begin{equation}\label{eq35} 
\begin{array}{l}
G_{\alpha \beta} = 
\begin{pmatrix}
0 & x_+x_- & 0 \\  x_+x_- & 0 & 0 \\ 0 & 0 & -r^2
\end{pmatrix}\\
G^{\alpha \beta} = 
\begin{pmatrix}
0 & \frac{1}{x_+x_-} & 0 \\  \frac{1}{x_+x_-} & 0 & 0 \\ 0 & 0 & -\frac{1}{r^2}
\end{pmatrix}\\
\sqrt{G} = r|x_+x_-|.
\end{array}
\end{equation}
%%%%%%%%%%%%%%%%%%%%%%%%%%%%%%%%%%%%%%%%%%%%%%%%%%%%%%%%%%%%%%%%%%%%%%%%%%%%%
%PAGE 11 PAGE 11 PAGE 11 PAGE 11 PAGE 11 PAGE 11 PAGE 11 PAGE 11 PAGE 11 PAGE 11 
%%%%%%%%%%%%%%%%%%%%%%%%%%%%%%%%%%%%%%%%%%%%%%%%%%%%%%%%%%%%%%%%%%%%%%%%%%%%% 
\begin{equation}\label{eq36} 
\partial_{\alpha}\sqrt{G}G^{\alpha\beta}\partial_{\beta} = \pm \bigg(\partial_+ r \partial_- + \partial_- r \partial_+ - \frac{x_+x_-}{r}\partial^2_{\theta} \bigg),
\end{equation}
applied to $x^{\mu}$, gives \eqref{eq32}-\eqref{eq34} ($\pm$ corresponding to the sign of $x_+x_-$).
Note the slight, but crucial, difference to the case of a
string world-sheet, where \eqref{eq31} implies $\sqrt{G}G^{\alpha \beta} = \pm
\big(
\begin{smallmatrix}
0 &1 \\ 1& 0
\end{smallmatrix}
\big)
$, hence minimal surfaces in 3-dimensional Minkowski-space corresponding to sums of 2 Null-curves ($x = \psi(\theta_+) + \phi(\theta_-)$, $\psi'^2 = 0 = \phi'^2$). For axially symmetric membranes \eqref{eq35} results in equations whose solutions due to the extra factor(s) of $r$, are `not exactly' sums of two Null-curves, but `rather close'. Given the crucial conditions \eqref{eq31} each of the second order equations will imply the other ones. Note that this formulation is manifestly invariant with respect to boosts in the $z$-direction, i.e. `hyperbolic rotations'
\begin{equation}\label{eq37} 
\begin{split}
\tilde{t} & = c_\gamma t + s_\gamma z \\
\tilde{z} & = s_\gamma t + c_\gamma z. 
\end{split}
\end{equation}
%%%%%%%%%%%%%%%%%%%%%%%%%%%%%%%%%%%%%%%%%%%%%%%%%%%%%%%%%%%%%%%%%%%%%%%%%%%%%
%PAGE 12 PAGE 12 PAGE 12 PAGE 12 PAGE 12 PAGE 12 PAGE 12 PAGE 12 PAGE 12 PAGE 12 
%%%%%%%%%%%%%%%%%%%%%%%%%%%%%%%%%%%%%%%%%%%%%%%%%%%%%%%%%%%%%%%%%%%%%%%%%%%%% 
Consider now what happens to \eqref{eq31}/\eqref{eq32} when transforming to the $(u,v)$ (or $(t,\varphi)$) coordinates that have been discussed above. For $(\theta_+\theta_-) \leftrightarrow (u(= z(\theta_+, \theta_-)), v)$ e.g.
\begin{equation}\label{eq38} 
\begin{split}
\bigg( \frac{\partial uv}{\partial\theta_+ \partial\theta_-} \bigg) & = 
\begin{pmatrix}
u_+ & u_- \\v_+ & v_-
\end{pmatrix} \\ 
& =
\begin{pmatrix} 
\partial_u \theta_+ &\partial_v \theta_+ \\ \partial_u \theta_- & \partial_v \theta_-
\end{pmatrix}^{-1} \\
& =
\frac{1}{\Delta}
\begin{pmatrix}
\partial_v \theta_- & -\partial_v \theta_+ \\ -\partial_u \theta_- & \partial_u \theta_+
\end{pmatrix}
\end{split}
\end{equation} 
one gets
\begin{equation}\label{eq39} 
\begin{split}
0 & = t^2_{\pm} - r^2_{\pm} - z^2_{\pm} \\
& = (T_u u_{\pm} + T_v v_{\pm})^2 - (R_u u_{\pm} + R_v v_{\pm})^2 -u_{\pm}^2 \\
& = u_{\pm}^2(T^2_u-R^2_u-1) - v_{\pm}^2(R^2_v-T^2_v) + 2u_{\pm}v_{\pm}(T_uT_v-R_uR_v)\\
& = (R^2_v-T^2_v)(u^2_{\pm}R^2-v^2_{\pm}) \\
& \stackrel{!}{=} 0,
\end{split}
\end{equation}
i.e.
\begin{equation}\label{eq40} 
u_{\pm}R = \mp v_{\pm},
\end{equation}
corresponding to $\Delta > 0$; equal signs in \eqref{eq40}, allowed for by \eqref{eq39}, must be excluded as otherwise the transformation between $\theta_+,\theta_-$ and $u,v$ could not be invertible.
Note that `again' the second order $z$-equation (now \eqref{eq33}) is automatic, as the compatibility condition following from \eqref{eq40}.
Using \eqref{eq38},
\begin{equation}\label{eq41} 
\partial_u\theta_{\pm} \pm R(u,v)\partial_v\theta_{\pm} = 0
\end{equation}
which from another point of view could also be taken as the definition of characteristic coordinates. 
\begin{equation}\label{eq42} 
\begin{split}
0 & = (T_u + RT_v)^2 - (R_u + RR_v)^2 - 1\\
  & = \big(t_+ (\partial_u + R\partial_v)\theta_+ + t_-(\partial_u + R\partial_v)\theta_-  \big)^2  \\
  &\hphantom{==} - \big(r_+ (\partial_u + R\partial_v)\theta_+ + (r_-\partial_u + R\partial_v)\theta_-  \big)^2-1\\
  & \stackrel{\eqref{eq41}}{=} 4(\partial_u\theta_-)^2(t^2_- - r^2_-) - 1 \\
  & = 4  (\partial_u\theta_-)^2 (t^2_- - r^2_- - z^2_-)
\end{split}
\end{equation}
(i.e \eqref{eq7} implying \eqref{eq30} when using \eqref{eq38}/\eqref{eq41}).
%%%%%%%%%%%%%%%%%%%%%%%%%%%%%%%%%%%%%%%%%%%%%%%%%%%%%%%%%%%%%%%%%%%%%%%%%%%%%
%PAGE 13 PAGE 13 PAGE 13 PAGE 13 PAGE 13 PAGE 13 PAGE 13 PAGE 13 PAGE 13 PAGE 13 
%%%%%%%%%%%%%%%%%%%%%%%%%%%%%%%%%%%%%%%%%%%%%%%%%%%%%%%%%%%%%%%%%%%%%%%%%%%%% 
\section{Symmetries}
\noindent
Although \eqref{eq1} and \eqref{eq7} at first sight appear to clearly not be invariant under the Lorentz-transformations \eqref{eq37}, they {\it are} if extending \eqref{eq37} to also involve the additional parameters, i.e. when involving explicit gauge-compensating transformations on $\varphi$, resp. $v$; namely when defining the following field-dependent transformations:
\begin{equation}\label{eq43} 
\begin{array}{l}
t=T(u,v)  \rightarrow \tilde{T}(u,v) = c_{\gamma} T(u,v) + s_{\gamma}u = c_{\gamma}t + s_{\gamma}z(t,\varphi)\\[0.15cm]
z(t,\varphi) = u  \rightarrow \tilde{u} = c_{\gamma} u + s_{\gamma}T(u,v) = c_{\gamma}z(t,\varphi) + s_{\gamma}t\\[0.15cm]
R^2_0v(t,\varphi)  \rightarrow R^2_0\tilde{v} = c_{\gamma}R^2_0 v + s_{\gamma}r^2_0 \phi(u,v) = c_{\gamma}R_0v(t,\varphi) + s_{\gamma}r^2_0(t,\varphi) \\[0.15cm]
r^2_0\varphi = r^2_0 \phi(u,v) \rightarrow r^2_0\tilde{\phi}(u,v) = c_{\gamma}r^2_0 \phi(u,v) + s_{\gamma}R^2_0v = c_{\gamma}r^2_0\varphi + s_{\gamma}R^2_0 v(t,\varphi)\\[0.15cm]
r = R \rightarrow \tilde{R}(u,v) = R(u,v) = r(t,\varphi) = \tilde{r}(t,\varphi) 
\end{array}
\end{equation}
(from now on again $r_0 = 1 = R_0$) where $\phi(u,v)$ (analogously for $v(t,\varphi)$), defined by \eqref{eq16}, in characteristic coordinates $(\theta_+,\theta_-)$, $\phi(u,v) = \phi(\theta_+,\theta_-)$ satisfies (in analogy to \eqref{eq40})
\begin{equation}\label{eq44} 
rt_{\pm} = \mp \phi_{\pm}
\end{equation}
(making \eqref{eq32} a compatibility consequence, resp. a consistent definition of $\varphi$ if \eqref{eq32} is satisfied) corresponding to 
\begin{equation}\label{eq45} 
\dot{\tilde{\theta}}_{\pm} \pm r(t,\varphi) \tilde{\theta}'_{\pm} = 0;
\end{equation}
here $\tilde{\theta}_{\pm}(t,\varphi) = \theta_{\pm}(u,v)$, and one can easily check the consistency of \eqref{eq45} with \eqref{eq41}:
\begin{equation}\label{eq46} 
\begin{split}
\dot{\tilde{\theta}}_{\pm}  \pm r\tilde{\theta}'_{\pm} & = \big( \dot{u}\partial_u\theta_{\pm} + \dot{v}\partial_v\theta_{\pm} \big)
\pm \big( ru'\partial_u\theta_{\pm} + v'r\partial_v\theta_{\pm} \big)\\
& = (\dot{u}-v')\partial_u\theta_{\pm} \pm \partial_u\theta_{\pm}(ru'-\frac{1}{r}\dot{v}) \\
& = 0.
\end{split}
\end{equation}
%%%%%%%%%%%%%%%%%%%%%%%%%%%%%%%%%%%%%%%%%%%%%%%%%%%%%%%%%%%%%%%%%%%%%%%%%%%%%
%PAGE 14 PAGE 14 PAGE 14 PAGE 14 PAGE 14 PAGE 14 PAGE 14 PAGE 14 PAGE 14 PAGE 14 
%%%%%%%%%%%%%%%%%%%%%%%%%%%%%%%%%%%%%%%%%%%%%%%%%%%%%%%%%%%%%%%%%%%%%%%%%%%%% 
One way to prove that \eqref{eq43} leaves \eqref{eq7} invariant is to first verify, by a slightly tedious (but straight-forward) calculation, using
\begin{equation}\label{eq47} 
\bigg( \frac{\partial uv}{\partial \tilde{u} \tilde{v}} \bigg) = \bigg( \frac{\partial \tilde{u} \tilde{v}}{\partial uv } \bigg)^{-1} = \frac{1}{\Delta}
\begin{pmatrix}
\tilde{v}_v & -\tilde{u}_v \\  -\tilde{v}_u & \tilde{u}_u
\end{pmatrix}
\end{equation}
\begin{equation}\label{eq48} 
\begin{split}
\partial_{\tilde{u}} & = \frac{1}{\Delta}\big( \tilde{v}_v\partial_u - \tilde{v}_u\partial_v \big)\\
\partial_{\tilde{v}} & = \frac{1}{\Delta}\big( -\tilde{u}_v\partial_u + \tilde{u}_u\partial_v \big)
\end{split}
\end{equation}
that \eqref{eq43} implies
\begin{equation}\label{eq49} 
\tilde{T}_{\tilde{u}}\tilde{T}_{\tilde{v}} = \tilde{R}_{\tilde{u}}\tilde{R}_{\tilde{v}}
\end{equation}
(both sides, multiplied by $\Delta^2$, turn out to be equal to $T_v\big( T_u(c_{\gamma}^2 + s^2_{\gamma}) + s_{\gamma}c_{\gamma}(1+T^2_u-R^2T^2_v)\big)$, when using \eqref{eq7} and \eqref{eq16}), and then note that (using \eqref{eq44}) \eqref{eq43} implies
\begin{equation}\label{eq50} 
\tilde{u}_{\pm}R = \mp\tilde{v}_{\pm},
\end{equation}
which (using an argument analogous to \eqref{eq39}) implies
\begin{equation}\label{eq51} 
\tilde{T}^2_u - \tilde{R}^2_u-1 = \tilde{R}^2(\tilde{R}^2_{\tilde{v}}-\tilde{T}^2_{\tilde{v}} ).
\end{equation}
%Mathematically, the significance of $r_0$ (resp. $R_0$) for compact membranes/ranges of the parameters/ is given by Moser's %lemma \cite{M65}. 
%%%%%%%%%%%%%%%%%%%%%%%%%%%%%%%%%%%%%%%%%%%%%%%%%%%%%%%%%%%%%%%%%%%%%%%%%%%%%
%PAGE 15 PAGE 15 PAGE 15 PAGE 15 PAGE 15 PAGE 15 PAGE 15 PAGE 15 PAGE 15 PAGE 15 
%%%%%%%%%%%%%%%%%%%%%%%%%%%%%%%%%%%%%%%%%%%%%%%%%%%%%%%%%%%%%%%%%%%%%%%%%%%%% 
\eqref{eq43} applied to any given solution (of \eqref{eq7}) will produce a(nother/new) solution; in the example $T = Hv$, $R = H\sqrt{v^2-1}$, $\phi = H'\frac{1}{2}(v^2-1)$ e.g. one would get $\tilde{T}(u,v) = cHv +su$, $\tilde{R} = R = H\sqrt{v^2-1}$, $\tilde{u} = sHv+cu$, $\tilde{v} = cv+sH'\frac{1}{2}(v^2-1)$.\\
%%%%%%%%%%%%%%%%%%%%%%%%%%%%%%%%%%%%%%%%%%%%%%%%%%%%%%%%%%%%%%%%%%%%%%%%%%%%%
%PAGE 15' PAGE 15' PAGE 15' PAGE 15' PAGE 15' PAGE 15' PAGE 15' PAGE 15' PAGE 15' 
%%%%%%%%%%%%%%%%%%%%%%%%%%%%%%%%%%%%%%%%%%%%%%%%%%%%%%%%%%%%%%%%%%%%%%%%%%%%% 
A second useful observation is to note that $\tilde{T}(\tilde{u}, \tilde{v}=v) = \lambda T\big(\frac{\tilde{u}}{\lambda},v \big)$, $\tilde{R}(\tilde{u}, \tilde{v}=v) = \lambda R\big(\frac{\tilde{u}}{\lambda},v \big)$, $\tilde{\phi}(\tilde{u}, \tilde{v}=v) = \phi\big(\frac{\tilde{u}}{\lambda},\tilde{v}=v \big)$, $\tilde{R}_0 = \lambda R_0$, corresponding to $\lambda \cdot \mathfrak{M}_3$, will solve the equations, if $T, R, \phi, R_0$ do. At this point some additional comments about $\tilde{\rho_0 = R_0^2}$ (the power $2$ having been chosen such that $R_0$ scales linearly with $\lambda$, s.a, resp. that with the Dirac-solution, $\dot{r}^2(t)+\frac{r^4(t)}{r_0^4}=1$, $r_0$ corresponds to the maximal radius of the pulsating sphere; mathematically, the significance of $r_0$ (resp. $R_0$) for compact membranes/ranges of the parameters/ is given by Moser's lemma \cite{M65}) are perhaps useful: first of all note that the above $\lambda$-scaling is a `physical/geometric' scaling, meaning: would in the description $2$ similarly describe $\lambda \mathfrak{M}_3$, and in particular require $\tilde{r}_0 = \lambda r_0$. 
Secondly, two reasons for not always having put $\tilde{\rho}_0$ (or/and $\rho_0$) $=1$ (which would have saved one with carrying factors of $R_0$ and $r_0$ around): if using the above scaling symmetry to reach the value $1$, which on either one side one could certainly do, it would a priori not be clear (nor, most likely, be the case) whether that particular scaling would make the conserved constant $=1$ on the other side; if on the other hand, using $v\rightarrow \kappa v$ and or scaling the `angle' $\varphi$ (on each side {\it those} scalings {\it could} be used independently) this could for compact membranes cause multi-coverings; in examples with non-compact range, like the example \cite{H22} it would be possible and convenient to put $\rho=1$. \\
%%%%%%%%%%%%%%%%%%%%%%%%%%%%%%%%%%%%%%%%%%%%%%%%%%%%%%%%%%%%%%%%%%%%%%%%%%%%%
%PAGE 15' PAGE 15' PAGE 15' PAGE 15' PAGE 15' PAGE 15' PAGE 15' PAGE 15' PAGE 15' 
%%%%%%%%%%%%%%%%%%%%%%%%%%%%%%%%%%%%%%%%%%%%%%%%%%%%%%%%%%%%%%%%%%%%%%%%%%%%% 
Both symmetries (boosts, and scaling) however, though on each side differently implemented (in the $(u,v)$ resp.$(t,\varphi)$ parametrization) have the same geometric action/effect on the two sides (in case of the scaling symmetry e.g. describing `$\lambda\cdot\mathfrak{M}^3$' instead of $\mathfrak{M}_3$). Still, the existence, (and simplicity) of the (`B\"acklund') transformation \eqref{eq16} that maps solutions in one parametrization to solutions in the other {\it is} a sign of `integrability'.\\
%%%%%%%%%%%%%%%%%%%%%%%%%%%%%%%%%%%%%%%%%%%%%%%%%%%%%%%%%%%%%%%%%%%%%%%%%%%%%
%PAGE 16 PAGE 16 PAGE 16 PAGE 16 PAGE 16 PAGE 16 PAGE 16 PAGE 16 PAGE 16 PAGE 16 
%%%%%%%%%%%%%%%%%%%%%%%%%%%%%%%%%%%%%%%%%%%%%%%%%%%%%%%%%%%%%%%%%%%%%%%%%%%%% 
Note that, curiously, each of the equations in \eqref{eq22}, {\it alone}, leads to the equation
\begin{equation}\label{eq52} 
\begin{split}
\tilde{D}_{\pm}[R] & := R_{uu} - R^2R_{vv} - RR^2_v - RT^2_v\\
& \big(= \tilde{D}_0[R] - \frac{1}{R}F^2_{\mp}\; \big) = 0,
\end{split}
\end{equation}
if $F$ in \eqref{eq22} is chosen to be $F_+$ or $F_-$, 
\begin{equation}\label{eq53} 
F_{\pm} = \frac{1}{2} \big( \sqrt{1+(R_u+RR_v)^2} \pm \sqrt{1+(R_u-RR_v)^2}\, \big),
\end{equation}
while the fact that (due to the relation of $T$ to $\phi$, resp. the PDE satisfied by $T$) {\it both} equations in \eqref{eq22} hold can be used to derive that
\begin{equation}\label{eq54} 
\begin{pmatrix}
F_u \\F_v
\end{pmatrix} = \frac{1}{1-\frac{Q^2}{F^4}}
\begin{pmatrix}
1 & -\frac{RQ}{F^2} \\
-\frac{Q}{RF^2} & 1
\end{pmatrix} 
\frac{1}{F}
\begin{pmatrix}
(RQ)_v \\(\frac{Q}{R})_u
\end{pmatrix}
\end{equation}
where $Q := RR_uR_v$.
%%%%%%%%%%%%%%%%%%%%%%%%%%%%%%%%%%%%%%%%%%%%%%%%%%%%%%%%%%%%%%%%%%%%%%%%%%%%%
%PAGE 17 PAGE 17 PAGE 17 PAGE 17 PAGE 17 PAGE 17 PAGE 17 PAGE 17 PAGE 17 PAGE 17  
%%%%%%%%%%%%%%%%%%%%%%%%%%%%%%%%%%%%%%%%%%%%%%%%%%%%%%%%%%%%%%%%%%%%%%%%%%%%% 
\section{Light-Cone Formulation}
\noindent
Yet another triply orthogonal parametrization (given already in \cite{H82}, but rarely used -- see however p.52-55 of \cite{H19}) should be mentioned, namely
\begin{equation}\label{eq55} 
\begin{split}
\hat{x}^{\mu} & = 
\begin{pmatrix}
\tau + \frac{\zeta}{2} \\[0.15cm]  R{\cos v \atop \sin v} \\[0.15cm] \tau-\frac{\zeta}{2}
\end{pmatrix}, \quad
\partial_{\tau}x^{\mu} = 
\begin{pmatrix}
1 + \frac{\dot{\zeta}}{2} \\[0.15cm]  \dot{R}{\cos v \atop \sin v} \\[0.15cm] 1 - \frac{\dot{\zeta}}{2}
\end{pmatrix}, \\
x'^{\mu} & = 
\begin{pmatrix}
\frac{\zeta'}{2} \\[0.15cm]  R'{\cos v \atop \sin v} \\[0.15cm] -\frac{\zeta'}{2}
\end{pmatrix}, \quad
\hat{G}_{\alpha \beta}  = 
\begin{pmatrix}
2\dot{\zeta} - \dot{R}^2 & 0 & 0 \\ 0 & -R'^2 & 0 \\ 0 & 0 & -R^2
\end{pmatrix}
\end{split}
\end{equation}
with
\begin{equation}\label{eq56} 
\big( \hat{G}_{01} =  \big)\zeta' - \dot{R}R' = 0, \quad 2\dot{\zeta} = \dot{R}^2 + \frac{R^2R'^2}{\eta^2}, 
\end{equation}
i.e. $\hat{G}_{01} = 0\,(\equiv\hat{G}_{10})$ and $\hat{G}_{11}\hat{G}_{22} = \eta^2\hat{G}_{00}$
($R$ and $\zeta$ are functions of $\tau$ and a spatial parameter, $\mu$). \eqref{eq56} implies {\it both}
\begin{equation}\label{eq57} 
\eta^2 \ddot{R} = R^2R'' + RR'^2 \,\big(=R(RR')' = (R^2R')'-R'R^2 \big),
\end{equation} 
and 
\begin{equation}\label{eq58} 
\eta^2\ddot{\zeta} = (R^2 \zeta')',
\end{equation} 
which are the correct equations,
\begin{equation}\label{eq59} 
\partial_{\alpha}\sqrt{\hat{G}}\hat{G}^{\alpha\beta}\partial_{\beta}\hat{x}^{\mu} =
\big( \partial^2_{\tau} - \frac{1}{\eta^2}\partial_{\mu}R^2\partial_{\mu} - \frac{R'^2}{\eta^2}\partial_v^2 \big)\hat{x}^{\mu} = 0.
\end{equation} 
While the fact that \eqref{eq57} are {\it Hamiltonian} equations, with respect to
\begin{equation}\label{eq60} 
H[R, P; \eta, \zeta_0] = \frac{1}{2\eta} \int (P^2 + R^2R'^2) d\mu
\end{equation} 
(with the weight, $\eta = \int \rho_- (\mu) d\mu$, of the conserved light-cone density $\rho_-$acquiring yet another meaning/importance)
%%%%%%%%%%%%%%%%%%%%%%%%%%%%%%%%%%%%%%%%%%%%%%%%%%%%%%%%%%%%%%%%%%%%%%%%%%%%%
%PAGE 18 PAGE 18 PAGE 18 PAGE 18 PAGE 18 PAGE 18 PAGE 18 PAGE 18 PAGE 18 PAGE 18 
%%%%%%%%%%%%%%%%%%%%%%%%%%%%%%%%%%%%%%%%%%%%%%%%%%%%%%%%%%%%%%%%%%%%%%%%%%%%% 
is well-noted, it is in the context of integrability most likely crucial to {\it not} `forget' $\zeta$ (the traditional point of view has been to view \eqref{eq56} solely as determining $\zeta$, which-when considering \eqref{eq60} -- has `dropped out', except for the zero-mode $\zeta_0$, canonically conjugate to $\eta$; it should be worthwhile studying in detail the `reconstruction algebra' , introduced in \cite{H11}, for this axially symmetric case). That \eqref{eq56} equally implies \eqref{eq58} could be used to introduce $Y(\tau, \mu)$ via
\begin{equation}\label{eq61} 
\begin{array}{c}
(\eta^2 \dot{R}^2 + R^2R'^2)\frac{1}{2}  = Y' \,(= \dot{\zeta})\\[0.20cm]
R^2\dot{R}R'  = \dot{Y} \, (=R^2\zeta'),
\end{array}
\end{equation}
implying
\begin{equation}\label{eq62} 
\big( \frac{1}{R^2}\dot{Y} \big)\dot{•} = Y'' \frac{1}{\eta^2}
\end{equation}
and then, with
\begin{equation}\label{eq63} 
w' := \zeta - \zeta_0, \quad \dot{w} = \frac{1}{\eta^2}Y
\end{equation}
get
\begin{equation}\label{eq64} 
\eta^2\ddot{w} = R^2w''.
\end{equation}
For one particular (`starting') solution $R$, could one take $w$ to be a {\it linear-combination} of specific solutions of \eqref{eq64}, which then (calculating $\zeta$ from \eqref{eq63}, and $R$ from \eqref{eq56}) would /ad infinitum/ give new solutions?).  
%%%%%%%%%%%%%%%%%%%%%%%%%%%%%%%%%%%%%%%%%%%%%%%%%%%%%%%%%%%%%%%%%%%%%%%%%%%%%
%PAGE 19 PAGE 19 PAGE 19 PAGE 19 PAGE 19 PAGE 19 PAGE 19 PAGE 19 PAGE 19 PAGE 19 
%%%%%%%%%%%%%%%%%%%%%%%%%%%%%%%%%%%%%%%%%%%%%%%%%%%%%%%%%%%%%%%%%%%%%%%%%%%%% 
Note also the following puzzle (and its resolution): under boosts in the $z$-direction, $\eta$ and $\tau$ are known to be multiplied by $e^{\gamma} \in \mathbb{R}$. How does that fit into the above mentioned relations with the ($t\varphi$) resp. ($u,v$) representation? (in particular: how can one justify $\mu$ in \eqref{eq60} to be {\it invariant} under such boosts?) 
\begin{equation}\label{eq65} 
\begin{pmatrix}
\overset{(\sim)}{G}_{00} & 0 & 0 \\ 0 & \overset{(\sim)}{G}_{11} & 0 \\ 0 & 0 & \overset{(\sim)}{G}_{22}
\end{pmatrix} =
\begin{pmatrix}
\dot{\tau} & \dot{\mu} & 0 \\ \tau' & \mu' & 0 \\ 0 & 0 & 1
\end{pmatrix}
\begin{pmatrix}
2\dot{\zeta}-\dot{R}^2 & 0 & 0 \\ 0 & -R'^2 & 0 \\ 0 & 0 & -R^2
\end{pmatrix}
\begin{pmatrix}
\dot{\tau} & \tau' & 0 \\ \dot{\mu} & \mu' & 0 \\ 0 & 0 & 1
\end{pmatrix}
\end{equation}
where $\dot{•}$ and $'$ denote differentiations with respect to $t$ and $\varphi$ ({\it or} $u$ and $v$, depending on what one takes on the left) always gives
\begin{equation}\label{eq66} 
\begin{array}{c}
\hat{G}_{00}\dot{\tau}\tau' + \hat{G}_{11}\dot{\mu}\mu' = 0, \quad \hat{\rho}^2 = \frac{\hat{G}_{11}\hat{G}_{22}}{\hat{G}_{00}} = \eta^2 = Const.,\\[0.25cm]
\dfrac{\dot{\mu}\mu'}{\dot{\tau}\tau'} = \dfrac{R^2}{\eta^2} = \dfrac{r^2}{\eta^2} = \dfrac{1}{\eta^2}\dfrac{\dot{\mu}\dot{\tau}}{\mu'\tau'}
\end{array}
\end{equation}
hence 
\begin{equation}\label{eq67} 
\eta\dot{\mu} = r^2 \tau', \quad \eta\mu' = \dot{\tau}.
\end{equation}
So
\begin{equation}\label{eq68} 
 2\eta\dot{\mu} = r^2z', \quad 2\eta\mu' = 1 + \dot{z},
\end{equation}
resp.
\begin{equation}\label{eq69} 
\begin{split}
2\eta \tilde{\mu}_u & = R^2  \tilde{\tau}(u,v) = R^2 T_v = \phi_u\\
2\eta \tilde{\mu}_v & = 1 + T_u = 1+\phi_v,
\end{split}
\end{equation}
hence
\begin{equation}\label{eq70} 
\begin{array}{l}
\tilde{\mu}(u,v)  = \big( v + \phi(u,v) \big)\frac{1}{2\eta}, \; \tilde{\tau}(u,v) = \frac{1}{2}\big(u + T(u,v)\big)\\[0.20cm]
2\eta \mu(t,\varphi) = \varphi + v(t,\varphi), \; 2\tau(t,\varphi) = t+z(t,\varphi)
\end{array}
\end{equation}
which indeed {\it is} invariant under \eqref{eq43}, as
\begin{equation}\label{eq71} 
v + \phi(u,v) \rightarrow (c_{\gamma} + s_{\gamma})\big(v+\phi(u,v)\big) = e^{\gamma}\big(v+\phi(u,v)\big)
\end{equation}
as well as making $\eta\frac{\partial}{\partial \tau}$ invariant too.
%%%%%%%%%%%%%%%%%%%%%%%%%%%%%%%%%%%%%%%%%%%%%%%%%%%%%%%%%%%%%%%%%%%%%%%%%%%%%
%PAGE 19' PAGE 19' PAGE 19' PAGE 19' PAGE 19' PAGE 19' PAGE 19' PAGE 19' PAGE 19' 
%%%%%%%%%%%%%%%%%%%%%%%%%%%%%%%%%%%%%%%%%%%%%%%%%%%%%%%%%%%%%%%%%%%%%%%%%%%%% 
Note that \eqref{eq60} and \eqref{eq70}/\eqref{eq16} imply (i.e. again, a conformal equivalence in the upper $2\times 2$ part of the metric)
\begin{equation}\label{eq72} 
\begin{split}
\frac{T^2_u-R^2_u-1}{[(1+T_u)^2 - R^2T_v^2]} & = \frac{s^2s'^2}{4\eta^2} \\[0.15cm]
\frac{R^2_v-T^2_v}{[(1+T_u)^2 - R^2T_v^2]} & = \frac{s'^2}{4\eta^2},
\end{split}
\end{equation}
which is consistent, as the ratio {\it is} $R^2$, and the second equation also follows by using
\begin{equation}\label{eq73} 
\begin{split}
s' = \partial_{\eta}s & = (u_{\eta}\partial_u + v_{\eta}\partial_v)R\\[0.15cm]
& = \frac{2\eta}{[(1+T_u)^2 - R^2T^2_v]}[(1+T_u)R_v-T_vR_u].
\end{split}
\end{equation}
Similarly, also using
\begin{equation}\label{eq74} 
\begin{split}
\zeta' & = (u_{\eta}\partial_u + v_{\eta}\partial_v)(T(u,v)-u) = T_v\frac{4\eta}{\Delta}\\
& \stackrel{!}{=} \dot{s}s'
\end{split}
\end{equation}
one finds
\begin{equation}\label{eq75} 
\begin{split}
\dot{s} & = \frac{2}{\Delta}\big( (1+T_u)\partial_u - R^2T_v\partial_v \big)R \\
& = \frac{2}{\Delta}\big( R_u(1+T_u)- R^2T_vR_v \big) \\
& = \frac{T_v}{\sqrt{\Delta}\sqrt{R^2_v-T^2_v}}
\end{split}
\end{equation}
and/hence
\begin{equation}\label{eq76} 
\big( (1+T_u)\partial_v - T_v\partial_u \big)\frac{T_v}{\sqrt{\Delta}\sqrt{R^2_v-T^2_v}} =
\big( (1+T_u)\partial_u - R^2T_v\partial_v \big)\frac{\sqrt{R^2_v-T^2_v} }{\sqrt{\Delta}}.
\end{equation}
%%%%%%%%%%%%%%%%%%%%%%%%%%%%%%%%%%%%%%%%%%%%%%%%%%%%%%%%%%%%%%%%%%%%%%%%%%%%%
%PAGE 20 PAGE 20 PAGE 20 PAGE 20 PAGE 20 PAGE 20 PAGE 20 PAGE 20 PAGE 20 PAGE 20 
%%%%%%%%%%%%%%%%%%%%%%%%%%%%%%%%%%%%%%%%%%%%%%%%%%%%%%%%%%%%%%%%%%%%%%%%%%%%% 
\section{More B\"acklund Transformations}
\noindent
Due to \eqref{eq16}, \eqref{eq23} implies {\it two} compatibility equations, namely (for $r_0 = 1 = R_0$)
\begin{equation}\label{eq77} 
[1]:= f'+ \big(\frac{1}{r}\tilde{f}\big)\dot{•} = 0, \quad [2]:= \dot{f} + (r\tilde{f})' = 0,  
\end{equation}
where $f$ is one of the 4 solutions
\begin{equation}\label{eq78} 
\begin{split}
f_{++} & = +\sqrt{p + \sqrt{p^2-q^2}} = +\sqrt{f^2_+},\\
f_{+-} & = -\sqrt{p + \sqrt{p^2-q^2}} = -\sqrt{f^2_+},\\
f_{-+} & = +\sqrt{f^2_-} = +\sqrt{p - \sqrt{p^2-q^2}},\\
f_{--} & = -\sqrt{p - \sqrt{p^2-q^2}}
\end{split}
\end{equation}
and $\tilde{f}$ another one, such that 
\begin{equation}\label{eq79} 
f\tilde{f} = r\dot{r}r' = q;
\end{equation}
each of the 4 solutions in \eqref{eq78} satisfies the `linear' PDE
\begin{equation}\label{eq80} 
(\frac{1}{r}\tilde{f} )\ddot{•} = ( r\tilde{f} )''
\end{equation}
and the set given in \eqref{eq78} coincides with the set (cp.\eqref{eq29})
\begin{equation}\label{eq81} 
g_{\delta\delta'} = \frac{\delta}{2}\sqrt{1-(\dot{r}-rr')^2} + \frac{\delta'}{2}\sqrt{1-(\dot{r}+rr')^2} 
=: \frac{\delta}{2}\sqrt{-} +  \frac{\delta'}{2}\sqrt{+}.
\end{equation}
Using both parts of \eqref{eq77} somewhat simplifies the tedious derivation of \eqref{eq28}; 
e.g. for $f = g_{++} = f_+$, $\tilde{f} = g_{+-} = f_-$:
\begin{equation}\label{eq82} 
\begin{split}
r(\sqrt{-}' + \sqrt{+}') + (\sqrt{-} - \sqrt{+})\dot{•} - \frac{\dot{r}}{r}(\sqrt{-} - \sqrt{+}) & \stackrel{1}{=} 0\\
r(\sqrt{-}' - \sqrt{+}') + (\sqrt{-} + \sqrt{+})\dot{•} + r'(\sqrt{-} - \sqrt{+}) & \stackrel{2}{=} 0,
\end{split}
\end{equation}
$\frac{[1]+[2]}{\dot{r}-rr'}$ giving
\begin{equation}\label{eq83} 
(\partial_t + r\partial_{\varphi})(\partial_t - r\partial_{\varphi})r + \frac{1}{r}(p+q -\sqrt{p^2-q^2}) =0,
\end{equation}
%%%%%%%%%%%%%%%%%%%%%%%%%%%%%%%%%%%%%%%%%%%%%%%%%%%%%%%%%%%%%%%%%%%%%%%%%%%%%
%PAGE 21 PAGE 21 PAGE 21 PAGE 21 PAGE 21 PAGE 21 PAGE 21 PAGE 21 PAGE 21 PAGE 21 
%%%%%%%%%%%%%%%%%%%%%%%%%%%%%%%%%%%%%%%%%%%%%%%%%%%%%%%%%%%%%%%%%%%%%%%%%%%%% 
$\frac{[2]-[1]}{\dot{r}+rr'}$ giving
\begin{equation}\label{eq84} 
(\partial_t - r\partial_{\varphi})(\partial_t + r\partial_{\varphi})r + \frac{1}{r}(p-q -\sqrt{p^2-q^2}) =0,
\end{equation}
implying $\dot{r}\eqref{eq82}_1 = -rr'\eqref{eq82}_2$ -- note also (the overall sign in front of $\sqrt{\hphantom{x}}$ being positive if $f = f_+$)
\begin{equation}\label{eq85} 
(\partial_t \pm r\partial_{\varphi})z = f \mp \tilde{f} = \sqrt{\pm} = \sqrt{1-(\dot{r} \pm rr')^2}
\end{equation}
and that \eqref{eq82} implies
\begin{equation}\label{eq86} 
(\partial_t \pm r\partial_{\varphi})\sqrt{\mp} = \pm \frac{1}{2r}(\dot{r} \mp rr')(\sqrt{-} - \sqrt{+});
\end{equation}
and that the two equations in \eqref{eq82} (resp. \eqref{eq77}) in characteristic coordinates read (consistent with \cite{H19}; note that $2\dot{r} = \frac{r_+}{t_+} + \frac{r_-}{t_-}$)
\begin{equation}\label{eq87} 
\big( \frac{t_{+ -}}{t_+t_-} +\frac{\dot{r}}{r} \big) \cdot  \big( \frac{z_+}{t_+} - \frac{z_-}{t_-} \big) = 0
\end{equation}
resp.
\begin{equation}\label{eq88} 
2z_{+-} = \frac{z_-}{t_-} (t_{+-} - r't_+t_-) + \frac{z_+}{t_+} (t_{+-} + r't_+t_-),
\end{equation}
but it is important to stress that, as done when first deriving \eqref{eq28} (= \eqref{eq84} = \eqref{eq83}), each of the two equations in \eqref{eq77}, {\it alone}, is sufficient to give \eqref{eq28}, -- whose solutions, via \eqref{eq23} and \eqref{eq2}, give (not counting $z \rightarrow -z$) `half' of the (time-like) axially minimal 3-manifolds in $\mathbb{R}^{1,3}$. This may lead one to speculate that the 2 (equivalent) options in \eqref{eq77} reflect a bi-Hamiltonian nature of the problem (which is one of the formal routes to get infinitely many conserved quantities); but where (what exactly) are the Hamiltonians?\\
%%%%%%%%%%%%%%%%%%%%%%%%%%%%%%%%%%%%%%%%%%%%%%%%%%%%%%%%%%%%%%%%%%%%%%%%%%%%%
%PAGE 22 PAGE 22 PAGE 22 PAGE 22 PAGE 22 PAGE 22 PAGE 22 PAGE 22 PAGE 22 PAGE 22 
%%%%%%%%%%%%%%%%%%%%%%%%%%%%%%%%%%%%%%%%%%%%%%%%%%%%%%%%%%%%%%%%%%%%%%%%%%%%% 
First of all, what about a {\it Lagrangian} description of \eqref{eq28}? One could try to `parametrize' the unparametric radial action for $r = r(t,z)$ (cp. e.g. \cite{H14})
\begin{equation}\label{eq89} 
S_r[r(t,z)] = \int r \sqrt{1-\dot{r}^2 + r'^2} dt dz,
\end{equation}
i.e. considering $(t = x^0, \, x^3 = z(t,\varphi)) \leftrightarrow (t = \varphi^0, \, \varphi = \varphi^1(x^0, x^3)) $,
\begin{equation}\label{eq90} 
\begin{array}{l}
\bigg(\frac{\partial \varphi}{\partial x}  \bigg) = 
\begin{pmatrix}
1 & 0 \\ \partial_0\varphi & \partial_3\varphi
\end{pmatrix} =
\begin{pmatrix}
1 & 0 \\ \dot{z} & z'
\end{pmatrix}^{-1} =
\begin{pmatrix}
1 & 0 \\ -\frac{\dot{z}}{z'} & \frac{1}{z'}
\end{pmatrix}\\[0.45cm]
(\dot{r} =)\partial_{x^0}r = (\partial_t + \frac{\partial \varphi}{\partial x^0}\partial_{\varphi})r = 
(\partial_t - \frac{\dot{z}}{z'}\partial_{\varphi})r(\rightarrow \dot{r} + \frac{f}{\tilde{f}}rr')\\[0.35cm]
(r' =)\partial_{x^3}r = \frac{\partial \varphi}{\partial x^3}\partial_{\varphi}r = 
\frac{1}{z'}\partial_{\varphi} r (\rightarrow - \frac{rr'}{\tilde{f}})
\end{array}
\end{equation}
which (using $dt dz = dt d\varphi\vert \frac{\partial x^{\circ}}{\partial \varphi^{\circ}}\vert = |z'| dt d\varphi$) gives
\begin{equation}\label{eq91} 
\begin{split}
S_{rz} & =   S[r(t,\varphi), z(t,\varphi)] = \int r \sqrt{r'^2 + z'^2 - (\dot{r}z' -r'\dot{z})^2}\, dt d\varphi\\
 & \big( \rightarrow S_r [r(t,\varphi)]= \int \sqrt{r^2 r'^2 + \tilde{f}^2 - (\dot{r}\tilde{f} + rr'f)^2}\, dt d\varphi \, ?\big),
\end{split}
\end{equation}
with $f$, $\tilde{f}$ given by \eqref{eq29}. \\
While $S_{rz}$ is certainly correct (and coincides with $\int \sqrt{G} d\varphi^0 d\varphi d\psi$ when $\varphi^0 = t$, and integrating out the axial symmetry variable $\psi$, $S_r[r(t,\varphi)]$ must be taken with great care, as inserting \eqref{eq23} simply into the Lagrangian, rather than into the equations of motion could easily give wrong conclusions). 
%%%%%%%%%%%%%%%%%%%%%%%%%%%%%%%%%%%%%%%%%%%%%%%%%%%%%%%%%%%%%%%%%%%%%%%%%%%%%
%PAGE 23 PAGE 23 PAGE 23 PAGE 23 PAGE 23 PAGE 23 PAGE 23 PAGE 23 PAGE 23 PAGE 23 
%%%%%%%%%%%%%%%%%%%%%%%%%%%%%%%%%%%%%%%%%%%%%%%%%%%%%%%%%%%%%%%%%%%%%%%%%%%%% 
As $S_{rz}$ is unconstrained, one may wonder how it relates to \eqref{eq1}, whose first part is easily motivated/`achieved' by noting that \eqref{eq91} is invariant under time-dependent reparametrizations $\varphi \rightarrow \tilde{\varphi}(\varphi,t)$, which clearly allows to choose $\dot{r}r' + \dot{z}z' = 0$. 
Note that defining $\Pi_r := \frac{\delta \mathcal{L}}{\delta \dot{r}}$ and  $\Pi_z := \frac{\delta \mathcal{L}}{\delta \dot{z}}$ does give nice expression in terms of the Jacobian $(\dot{r}z' - r'\dot{z})$, and in particular $\Pi_r\dot{r} + \Pi_z\dot{z} = 0$, but -- no surprise -- does not allow to express $\dot{r}$ and $\dot{z}$ in terms of $\Pi_r$ and $\Pi_z$.\\
More importantly: what is the `symmetry' allowing for the {\it second} part of \eqref{eq1}?
That one can indeed choose $\varphi(r,z)$ such that {\it both parts} in \eqref{eq1} hold is specific to the minimal-surface problem, namely $t(r,z)$ (the time at which the surface reaches the point $(r,z)$ in space) satisfying (cp.\cite{H14})
\begin{equation}\label{eq92} 
\vec{\nabla}\bigg( \frac{r\cdot \vec{\nabla t}}{\sqrt{(\nabla t)^2 -1}} \bigg) = 0.
\end{equation}
%%%%%%%%%%%%%%%%%%%%%%%%%%%%%%%%%%%%%%%%%%%%%%%%%%%%%%%%%%%%%%%%%%%%%%%%%%%%%
%PAGE 24 PAGE 24 PAGE 24 PAGE 24 PAGE 24 PAGE 24 PAGE 24 PAGE 24 PAGE 24 PAGE 24 
%%%%%%%%%%%%%%%%%%%%%%%%%%%%%%%%%%%%%%%%%%%%%%%%%%%%%%%%%%%%%%%%%%%%%%%%%%%%% 
After the hodograph transformation $t, \varphi \leftrightarrow r,z$ (cp.\cite{H14},\cite{H19}), 
\begin{equation}\label{eq93} 
\begin{pmatrix}
\dot{r} & r' \\ \dot{z} & z'
\end{pmatrix} =
\begin{pmatrix}
t_r & t_z \\ \varphi_r & \varphi_z
\end{pmatrix}^{-1} = 
\frac{1}{det}
\begin{pmatrix}
\varphi_z & -t_z \\ -\varphi_r & t_r
\end{pmatrix}
\end{equation}
\eqref{eq1} reads
\begin{equation}\label{eq94} 
\vec{\nabla}\varphi \cdot \vec{\nabla}t = 0, \quad \big( (\nabla t)^2 -1 \big)\big( (\nabla \varphi)^2 - r^2 \big) =r^2,
\end{equation}
which {\it is} solvable, resp. solved, by
\begin{equation}\label{eq95} 
\varphi_r = \frac{\pm r t_z}{\sqrt{(\nabla t)^2 -1}}, \quad \varphi_z = \frac{\mp r t_r}{\sqrt{(\nabla t)^2 -1}}
\end{equation}
(the consistency being precisely \eqref{eq92}); as on the other hand
\begin{equation}\label{eq96} 
t_r = \frac{\mp \varphi_z}{\sqrt{(\nabla \varphi)^2 -r^2}}, \quad t_z = \frac{\pm \varphi_r}{\sqrt{(\nabla \varphi)^2 -r^2}},
\end{equation}
we have derived yet another sign of integrability, namely \eqref{eq95}/\eqref{eq96} being classical B\"acklund-transformations between solutions of \eqref{eq92} and solutions of 
\begin{equation}\label{eq97} 
\bigg( \frac{\varphi_r}{\sqrt{(\nabla \varphi)^2 -r^2}} \bigg)_r +
\bigg( \frac{\varphi_z}{\sqrt{(\nabla \varphi)^2 -r^2}} \bigg)_z =
\vec{\nabla} \bigg( \frac{\vec{\nabla}\varphi}{\sqrt{(\nabla \varphi)^2 -r^2}} \bigg) = 0.
\end{equation}
%%%%%%%%%%%%%%%%%%%%%%%%%%%%%%%%%%%%%%%%%%%%%%%%%%%%%%%%%%%%%%%%%%%%%%%%%%%%%
%PAGE 25 PAGE 25 PAGE 25 PAGE 25 PAGE 25 PAGE 25 PAGE 25 PAGE 25 PAGE 25 PAGE 25 
%%%%%%%%%%%%%%%%%%%%%%%%%%%%%%%%%%%%%%%%%%%%%%%%%%%%%%%%%%%%%%%%%%%%%%%%%%%%%
In analogy with \eqref{eq91} one could of course equally well, via $(t =T,\, z=u) \leftrightarrow (u,\, v= V(u, T(u,v))$, derive
\begin{equation}\label{eq98} 
S_{TR} = \int \sqrt{(R_uT_v - R_vT_u)^2 - (R^2_v - T^2_v)}\,R\, dudv
\end{equation}
from \eqref{eq89} (or \eqref{eq5}) and then, by choosing $v =V(u, T(u,v))$ such that \eqref{eq7} (implying \eqref{eq8}) holds, and again interchanging independent and dependent variables $(u,v)\leftrightarrow T, R,\, u =z(T,R),\,v=v(T,R)$,
\begin{equation}\label{eq99} 
\begin{pmatrix}
T_u & T_v \\ R_u & R_v
\end{pmatrix} =
\begin{pmatrix}
u_T & u_R \\ v_T & v_R
\end{pmatrix}^{-1} =
\frac{1}{\det}
\begin{pmatrix}
v_R & -u_R \\ -v_T & u_T
\end{pmatrix},
\end{equation}
obtaining
\begin{equation}\label{eq100} 
\begin{array}{c}
u_Tv_T = u_Rv_R \;( i.e. \partial_{\alpha}u \partial^{\alpha}v = 0)\\[0.25cm]
v^2_R-v^2_T-(u_Tv_R - u_Rv_T)^2 = R^2(u^2_T-u^2_R) \;(=R^2u^{\alpha}u_{\alpha})\\[0.25cm]
(1-u^{\alpha}u_{\alpha})(R^2-v^{\beta}v_{\beta}) = R^2,
\end{array}
\end{equation}
which is solved by
\begin{equation}\label{eq101} 
\begin{split}
v_T & = \frac{Ru_R}{\sqrt{1-u^{\alpha}u_{\alpha}}}, \; v_R  = \frac{Ru_T}{\sqrt{1-u^{\alpha}u_{\alpha}}},\\
u_R & = \frac{v_T}{\sqrt{R^2-v^{\beta}v_{\beta}}}, \; u_T  = \frac{v_R}{\sqrt{R^2-v^{\beta}v_{\beta}}},
\end{split}
\end{equation}
implying
\begin{equation}\label{eq102} 
\partial_{\alpha}\bigg( \frac{rz^{\alpha}}{\sqrt{1-z^{\beta}z_{\beta}}} \bigg) = 0, \;
\partial_{\alpha}\bigg( \frac{v^{\alpha}}{\sqrt{R^2-v^{\beta}v_{\beta}}} \bigg) = 0
\end{equation}
(which {\it are} the Euler-Lagrange equations corresponding to $-\int r\sqrt{1-\dot{z}^2 + z'^2}\,drdt$, resp.  $-\int \sqrt{R^2-\dot{v}^2 + v'^2}\,dRdt$).
%%%%%%%%%%%%%%%%%%%%%%%%%%%%%%%%%%%%%%%%%%%%%%%%%%%%%%%%%%%%%%%%%%%%%%%%%%%%%
%PAGE 26 PAGE 26 PAGE 26 PAGE 26 PAGE 26 PAGE 26 PAGE 26 PAGE 26 PAGE 26 PAGE 26 
%%%%%%%%%%%%%%%%%%%%%%%%%%%%%%%%%%%%%%%%%%%%%%%%%%%%%%%%%%%%%%%%%%%%%%%%%%%%%
\section{Zero (Gauss) Curvature Condition(s) for Unconstrained Motion(s)}
\noindent
The (semi-) final, calculationally (together with \eqref{eq28}) most difficult/tedious (though beautiful,as giving the simplest possible of all zero-curvature conditions), aspect of integrability in the extremality-properties of axially symmetric membranes, reported here, has to do with the reduced description following from \eqref{eq23}, namely considering simply the planar motion of the curves
\begin{equation}\label{eq103} 
\begin{split}
\vec{u}(t, \varphi) & = 
\begin{pmatrix}
r(t, \varphi) \\ z(t, \varphi)
\end{pmatrix}\\
\dot{\vec{u}}(t, \varphi) & = 
\begin{pmatrix}
\dot{r}\\ \dot{z}
\end{pmatrix} = 
\begin{pmatrix}
\dot{r}\\ f[r]
\end{pmatrix}\\
\vec{u}'(t, \varphi) & = 
\begin{pmatrix}
r'\\ z'
\end{pmatrix} = 
\begin{pmatrix}
r'\\ \frac{-\tilde{f}[r]}{r}
\end{pmatrix}
\end{split}
\end{equation}
resp.
\begin{equation}\label{eq104} 
\begin{split}
\vec{v}(t, \varphi) & = 
\begin{pmatrix}
r(t, \varphi) \\ v(t, \varphi)
\end{pmatrix}\\
\dot{\vec{v}}(t, \varphi) & = 
\begin{pmatrix}
\dot{r}\\ \dot{v}
\end{pmatrix} = 
\begin{pmatrix}
\dot{r}\\ -r\tilde{f}[r] = r^2z'
\end{pmatrix}\\
\vec{v}'(t, \varphi) & = 
\begin{pmatrix}
r'\\ v'
\end{pmatrix} = 
\begin{pmatrix}
r'\\ f[r] = \dot{z}
\end{pmatrix},
\end{split}
\end{equation}
which can be used to define 2-dimensional metrics
\begin{equation}\label{eq105} 
\begin{split}
(u_{ab}) & = 
\begin{pmatrix}
\dot{r}^2 + \dot{z}^2 = g_1 = H^2_1 = \dot{r}^2 + f^2 & 0 \\[0.15cm] 0 & r'^2 + z'^2 = g_2 = H^2_2 = r'^2 +  \frac{\tilde{f}^2}{r^2} 
\end{pmatrix}\\[0.25cm]
(\tilde{g}_{ab}) & = (v_{ab}) = 
\begin{pmatrix}
\dot{r}^2 + \dot{v}^2 = g_{11} = \dot{r}^2 + r^ 2\tilde{f}^2 & \dot{r}r' + \dot{v}v' = g_{12} = \dot{r}r' - r\tilde{f}f \\[0.15cm]
  & r'^2 + v'^2 = g_{22} = r'^2 + f^2
\end{pmatrix}
\end{split}
\end{equation}
%%%%%%%%%%%%%%%%%%%%%%%%%%%%%%%%%%%%%%%%%%%%%%%%%%%%%%%%%%%%%%%%%%%%%%%%%%%%%
%PAGE 27 PAGE 27 PAGE 27 PAGE 27 PAGE 27 PAGE 27 PAGE 27 PAGE 27 PAGE 27 PAGE 27 
%%%%%%%%%%%%%%%%%%%%%%%%%%%%%%%%%%%%%%%%%%%%%%%%%%%%%%%%%%%%%%%%%%%%%%%%%%%%%
(note that $(\tilde{g}_{ab}) = (v_{ab})$ is {\it not} diagonal, and its determinant is the square of $(\dot{r}\dot{z} - r^2r'z') = (\dot{r}f + rr'\tilde{f})$).
While calculating the Gauss-curvature $K$ from the LHS expressions in \eqref{eq105} one trivially gets zero (as the expressions for the entries of the metric then explicitly come from planar curves, sweeping out part of $\mathbb{R}^2$), this is {\it not} the case for the RHS expressions. For the simpler, diagonal, case Liouville's formula 
\begin{equation}\label{eq106} 
K = \frac{1}{\sqrt{g}}\left[ \big( \frac{\sqrt{g}}{g_{11}}\Gamma^2_{11} \big)' - \big( \frac{\sqrt{g}}{g_{11}}\Gamma^2_{12} \dot{\big)} \right] 
\end{equation}
simplifies to the standard consistency condition for diagonal metrics (see e.g. the first line of equation(3) in \cite{E1907},where, quoting Lam\'e, the conditions for triply orthogonal coordinate systems in $\mathbb{R}^3$ (!) are given),
\begin{equation}\label{eq107} 
\bigg(\frac{\dot{H_2}}{H_1} \dot{\bigg)} + \bigg(\frac{H'_1}{H_2} \bigg)' = 0 \, (\equiv -H_1H_2K),
\end{equation}
%%%%%%%%%%%%%%%%%%%%%%%%%%%%%%%%%%%%%%%%%%%%%%%%%%%%%%%%%%%%%%%%%%%%%%%%%%%%%
%PAGE 27' PAGE 27' PAGE 27' PAGE 27' PAGE 27' PAGE 27' PAGE 27' PAGE 27' PAGE 27' 
%%%%%%%%%%%%%%%%%%%%%%%%%%%%%%%%%%%%%%%%%%%%%%%%%%%%%%%%%%%%%%%%%%%%%%%%%%%%%
i.e.
\begin{equation}\label{eq108} 
\bigg( \frac{r'\dot{r}' + (\frac{\tilde{f}}{r})(\frac{\tilde{f}}{r})\dot{•}}{\dot{r}r'(\frac{f}{\dot{r}}+\frac{\dot{r}}{f})} \dot{\bigg)} + 
\bigg( \frac{\dot{r}\dot{r}' + ff'}{\dot{r}r'(\frac{f}{\dot{r}}+\frac{\dot{r}}{f})} \bigg)' = 0
\end{equation}
(note that they heavily involve $3^{\text{rd}}$-derivatives of $r$). Due to \eqref{eq23} one knows that, using the second-order equation for $r$, $D[r] = 0$ (cp. \eqref{eq28}), the above must hold, but it is important to find out `how exactly'. Using the first PDE in \eqref{eq77} (the one that was shown to reduce to \eqref{eq27}) one can write the two numerators above as 
\begin{equation}\label{eq109} 
\begin{split}
r'\dot{r}' + \big( \frac{\tilde{f}}{r} \big)& \big[ \big( \frac{\tilde{f}}{r} \dot{\big)} + f'\big]  - \dot{r}r'\frac{f'}{f} 
 = \frac{r'}{f} (\dot{r}' f - \dot{r}f') + \frac{\tilde{f}}{r} [1]\\
\dot{r}\dot{r}' + ff' & = \dot{r}\dot{r}' + f\big[f'+(\frac{\tilde{f}}{r})\dot{•} \big] - f\big(\frac{\dot{r}r'}{f} \dot{\big)}\\
& = - \ddot{r}r' + \dot{r}r' \frac{\dot{f}}{f} +f[1]\\
& = \frac{r'}{f}(-\ddot{r}f + \dot{r}\dot{f}) + f[1].
\end{split}
\end{equation}
%%%%%%%%%%%%%%%%%%%%%%%%%%%%%%%%%%%%%%%%%%%%%%%%%%%%%%%%%%%%%%%%%%%%%%%%%%%%%
%PAGE 28 PAGE 28 PAGE 28 PAGE 28 PAGE 28 PAGE 28 PAGE 28 PAGE 28 PAGE 28 PAGE 28  
%%%%%%%%%%%%%%%%%%%%%%%%%%%%%%%%%%%%%%%%%%%%%%%%%%%%%%%%%%%%%%%%%%%%%%%%%%%%%
As the (common) denominator contains only first derivatives, it is then easy to see that all third-order terms not involving [1] cancel, $(\ddot{r}'f - \dot{r}\dot{f}') + (\dot{r}\dot{f}' - \ddot{r}'f) = 0$, and the terms involving [1] give, with the help of \eqref{eq27} 
\begin{equation}\label{eq110} 
(aD)\dot{•} + (bD)' = \bigg( \frac{\dot{r}r'}{2f} \, \frac{D[r]}{\sqrt{p^2-q^2}} \dot{ \bigg)} + 
\bigg( f\frac{D[r]}{2\sqrt{p^2-q^2}} \bigg)' = 0   ,
\end{equation}
as the terms {\it not} involving [1] cancel 
\begin{equation}\label{eq111} 
\bigg( \frac{f\dot{r}' - f'\dot{r}}{f^2+\dot{r}^2}\dot{\bigg)}  + \bigg( \frac{\dot{f}\dot{r} - f\ddot{r}}{f^2+\dot{r}^2} \bigg)' \equiv 0
\end{equation}
(one way to see the cancellations in \eqref{eq111} is to look at the coefficients of $f,\,f^2\dot{f},\,f^2f',\,\dot{f},\,f',\,f\dot{f}f'$, i.e. {\it without} calculating the derivatives of $f$, or using its form; \eqref{eq111} holds for {\it any} $f$). Note that while $D[r] = 0$ implies $K=0$ which was clear from the beginning, the two conditions are not/`yet'/ equivalent, as \eqref{eq110} could in principle hold by a less trivial vanishing -mechanism. Where could additional `help' come from?  
%%%%%%%%%%%%%%%%%%%%%%%%%%%%%%%%%%%%%%%%%%%%%%%%%%%%%%%%%%%%%%%%%%%%%%%%%%%%%
%PAGE 29 PAGE 29 PAGE 29 PAGE 29 PAGE 29 PAGE 29 PAGE 29 PAGE 29 PAGE 29 PAGE 29  
%%%%%%%%%%%%%%%%%%%%%%%%%%%%%%%%%%%%%%%%%%%%%%%%%%%%%%%%%%%%%%%%%%%%%%%%%%%%%
To use \eqref{eq77}$_{[2]}$ for \eqref{eq108} seemed to not easily lead to conclusions. The $v$-curves, \eqref{eq104}, however (which by the central `B\"acklund' relations \eqref{eq16} also describe planar motion in terms of $f$ and $\tilde{f}$), though more complicated (s.b.), can be shown to give an equation of the form $(cD)\dot{•} + (dD)' = eD$, hence together with \eqref{eq110} presumably giving equivalence of $D[r] = 0$ and the vanishing of the Gauss curvature(s)
%$K=0$ 
(for motions satisfying \eqref{eq23}).\\ \eqref{eq106}, together with 
%%%%%%%%%%%%%%%%%%%%%%%%%%%%%%%%%%%%%%%%%%%%%%%%%%%%%%%%%%%%%%%%%%%%%%%%%%%%%
%PAGE 30 PAGE 30 PAGE 30 PAGE 30 PAGE 30 PAGE 30 PAGE 30 PAGE 30 PAGE 30 PAGE 30 
%%%%%%%%%%%%%%%%%%%%%%%%%%%%%%%%%%%%%%%%%%%%%%%%%%%%%%%%%%%%%%%%%%%%%%%%%%%%%
\begin{equation}\label{eq112} 
\begin{split}
\tilde{g}\Gamma^2_{11} & = -\frac{1}{2}g_{12}\dot{g}_{11} + \frac{1}{2}g_{11}(2\dot{g}_{12}-g'_{11})\\[0.15cm]
\tilde{g}\Gamma^2_{12} & = -\frac{1}{2}g_{12}g'_{11} + \frac{1}{2}g_{11}\dot{g}_{22}
\end{split}
\end{equation}
gives, with $\sqrt{\tilde{g}} = (f\dot{r}+ \tilde{f}rr')$,
\begin{equation}\label{eq113} 
\begin{split}
(f\dot{r}+ \tilde{f}rr') \tilde{K}  =  & \bigg[ \frac{(qr-\dot{r}r')(\dot{r}\ddot{r}+r\tilde{f}(r\tilde{f})\dot{•})}{(\dot{r}^2+\tilde{f}^2r^2)\sqrt{\tilde{g}}} + 
\frac{(\dot{r}r'-qr)\dot{•}-(\dot{r}\dot{r}'+\tilde{f}r(\tilde{f}r)'}{\sqrt{\tilde{g}}} \bigg]'\\[0.15cm]
& -\bigg[ \frac{r'\dot{r}' + f\dot{f}}{\sqrt{\tilde{g}}} + \frac{(qr-\dot{r}r')(\dot{r}\dot{r}'+ (\tilde{f}r)(\tilde{f}r)')}{\sqrt{\tilde{g}}(\dot{r}^2+\tilde{f}^2r^2)}\dot{\bigg]\hphantom{\big)}}. 
\end{split}
\end{equation}
Using \eqref{eq77} one may try to convince oneself that again all third-order terms {\it not} involving [1] or [2] cancel. As $D[r] = 0$ (which is of second-order) implies $\tilde{K} = 0$, \eqref{eq113} must then reduce to an equation of the form $(cD)\dot{•} + (dD)' = eD$, and it seems reasonable to assume that $D = 0$ is indeed equivalent to $K=0=\tilde{K}$.
%%%%%%%%%%%%%%%%%%%%%%%%%%%%%%%%%%%%%%%%%%%%%%%%%%%%%%%%%%%%%%%%%%%%%%%%%%%%%
%PAGE 32 PAGE 32 PAGE 32 PAGE 32 PAGE 32 PAGE 32 PAGE 32 PAGE 32 PAGE 32 PAGE 32 
%%%%%%%%%%%%%%%%%%%%%%%%%%%%%%%%%%%%%%%%%%%%%%%%%%%%%%%%%%%%%%%%%%%%%%%%%%%%%
\section{(Multi-) Hamiltonian Structures}
\noindent
The two equations in \eqref{eq102} are Hamiltonian with respect to 
\begin{equation}\label{eq114} 
\begin{split}
H_z & = \int dr \sqrt{\pi^2+r^2} \sqrt{1+z'^2} = H_z[z,\pi] \\
H_V & = \int dR \sqrt{P^2+1} \sqrt{R^2+V'^2} = H_V[V,P],
\end{split}
\end{equation}
as can be easily checked, using canonical Poisson structures:
\begin{equation}\label{eq115} 
\begin{split}
\dot{z} & = \frac{\delta H_z}{\delta \pi} = \pi \frac{\sqrt{1+z'^2}}{\sqrt{\pi^2+r^2}}, \quad
\dot{V}  = \frac{\delta H_V}{\delta P} = P \frac{\sqrt{R^2+V'^2}}{\sqrt{P^2+1}},\\[0.25cm]
\dot{\pi} & = -\frac{\delta H_z}{\delta z} = z' \frac{\sqrt{\pi^2+r^2}}{\sqrt{1+z'^2}}, \quad
\dot{P}  = -\frac{\delta H_V}{\delta V} = V' \frac{\sqrt{P^2+1}}{\sqrt{R^2+V'^2}}
\end{split}
\end{equation}
(the 2+1-dimensional version of $H_z$ was discussed in \cite{BH94}, and it's 1+1-dimensional version goes back to \cite{BC66}). Note also the relations  
\begin{equation}\label{eq116} 
\begin{split}
\dot{v} & = \frac{rz'}{\sqrt{1-\dot{z}^2 +z'^2}} = z'\frac{\sqrt{r^2+v'^2}}{\sqrt{1+z'^2}}, \\[0.15cm]
\dot{z} & = \frac{v'\sqrt{1+z'^2}}{\sqrt{r^2+v'^2}}, \quad
v'  = \frac{r\dot{z}}{\sqrt{1-\dot{z}^2+z'^2}}
\end{split}
\end{equation}
and
\begin{equation}\label{eq117} 
\bigg( {z' \atop v'}\dot{\bigg)} = \begin{pmatrix}
0 & \partial_r \\ \partial_r & 0
\end{pmatrix}
\begin{pmatrix}
\frac{\delta H}{\delta z'} \\[0.15cm] \frac{\delta H}{\delta v'}
\end{pmatrix}, \quad
H = \int \sqrt{1+z'^2} \sqrt{r^2+v'^2}\, dr.
\end{equation}
%%%%%%%%%%%%%%%%%%%%%%%%%%%%%%%%%%%%%%%%%%%%%%%%%%%%%%%%%%%%%%%%%%%%%%%%%%%%%
%PAGE 33 PAGE 33 PAGE 33 PAGE 33 PAGE 33 PAGE 33 PAGE 33 PAGE 33 PAGE 33 PAGE 33 
%%%%%%%%%%%%%%%%%%%%%%%%%%%%%%%%%%%%%%%%%%%%%%%%%%%%%%%%%%%%%%%%%%%%%%%%%%%%%
As $(q,p)\leftrightarrow (P = \mp q,\, Q = \mp p)$, i.e. interchanging coordinates and momenta (with one $-$ sign), are canonical transformations, from a Hamiltonian point of view the B\"acklund-transformations \eqref{eq101}, resp. \eqref{eq96}/\eqref{eq95}, could be considered as auto-B\"acklund-transformations ( as the Hamiltonians in \eqref{eq114} are `self-dual'). Also note that in the case of compact membranes, boundary conditions ( e.g. the range of $r$) are tacitly assumed to work out; e.g. in \eqref{eq117}, for the spherically symmetric solution,
\begin{equation}\label{eq118} 
\begin{array}{l}
z(t,r) = \sqrt{D^2(t)-r^2}, \quad \sqrt{1-\dot{z}^2+z'^2} = \frac{D^3}{\sqrt{D^2-r^2}}\\[0.25cm]
\dot{v} = \frac{-r^2}{D^3}, \quad v' = r\frac{\dot{D}}{D^2},\quad v(t,r) = \pm \frac{r^2}{2} \frac{\sqrt{1-D^4}}{D^2}, 
\end{array}
\end{equation}
\begin{equation}\label{eq119} 
\begin{split}
\frac{d}{dt}H & = \frac{d}{dt}\int_{0}^{D(t)} dr \sqrt{1 + z'^2}\, \sqrt{r^2+v'^2} \\
& = \frac{d}{dt} \lim_{\varepsilon \rightarrow 0} \int_{0}^{D(t)-\varepsilon} \frac{r\,dr}{D\sqrt{D^2-r^2}}\\
& = \frac{d}{dt} \lim_{\varepsilon \rightarrow 0}\bigg( \frac{-\sqrt{D^2-r^2}}{D}\bigg|_0^{D-\varepsilon} \bigg) =0 
\end{split}
\end{equation}
resp. (as one would do for other conserved quantities)
\begin{equation}\label{eq120} 
\frac{d}{dt}H = \lim_{\varepsilon \rightarrow 0}\bigg( \frac{\dot{D}}{\sqrt{2\varepsilon D}} - \frac{\dot{D}}{D^2} \int_0^{D-\varepsilon} \frac{r\,dr (2D^2-r^2)}{\sqrt{D^2-r^2}^3} \bigg) =  0
\end{equation}
%%%%%%%%%%%%%%%%%%%%%%%%%%%%%%%%%%%%%%%%%%%%%%%%%%%%%%%%%%%%%%%%%%%%%%%%%%%%%
%PAGE 34 PAGE 34 PAGE 34 PAGE 34 PAGE 34 PAGE 34 PAGE 34 PAGE 34 PAGE 34 PAGE 34 
%%%%%%%%%%%%%%%%%%%%%%%%%%%%%%%%%%%%%%%%%%%%%%%%%%%%%%%%%%%%%%%%%%%%%%%%%%%%%
\section{$D[r]$, Again}  
\noindent
Consider a conservation law
\begin{equation}\label{eq127} 
g'+\dot{h} = 0
\end{equation}
where $g=g(\dot{r},r',r)$ and $h(\dot{r},r',r)$ satisfy
\begin{equation}\label{eq128} 
g_u \equiv \frac{\partial g}{\partial \dot{r}} = -\frac{\partial h}{\partial r'} \equiv -h_w, \quad 
g_w \equiv \frac{\partial g}{\partial r'} = -r^2\frac{\partial h}{\partial \dot{r}} \equiv r^2h_u
\end{equation}
(which the 2 choices
$g = f_+ = \dot{z}_{(+)},\, h = \frac{1}{r}f_- = -z'_{(+)}$ resp.
$g = f_- = \dot{z}_{(-)},\, h = \frac{1}{r}f_+ = -z'_{(-)}$ both do) \eqref{eq127} then reads
\begin{equation}\label{eq129} 
h_{\dot{r}}(\ddot{r}-r^2r'') + \dot{r}h_r + r'g_r = 0;
\end{equation}
note that the terms proportional to $\dot{r}'$ have cancelled because of the $-$ sign in \eqref{eq128}$_1$ (in the more common, Lagrangian, origin of \eqref{eq127}, where $g = \frac{\partial \mathcal{L}}{\partial r'}$ and $h = \frac{\partial \mathcal{L}}{\partial \dot{r}}$ one would have a $+$ sign).\\
Assuming the 2 basic relations 
\begin{equation}\label{eq130} 
\begin{split}
g^2 + r^2h^2  & = 1 - \dot{r}^2 - r^2r'^2 \\
gh & = \dot{r}r'\; (=\frac{q}{r}), 
\end{split}
\end{equation}
(which actually {\it imply} \eqref{eq128}) one trivially derives
%%%%%%%%%%%%%%%%%%%%%%%%%%%%%%%%%%%%%%%%%%%%%%%%%%%%%%%%%%%%%%%%%%%%%%%%%%%%%
%PAGE 35 PAGE 35 PAGE 35 PAGE 35 PAGE 35 PAGE 35 PAGE 35 PAGE 35 PAGE 35 PAGE 35 
%%%%%%%%%%%%%%%%%%%%%%%%%%%%%%%%%%%%%%%%%%%%%%%%%%%%%%%%%%%%%%%%%%%%%%%%%%%%%
\begin{equation}\label{eq131} 
\begin{array}{l}
g_r h + gh_r  = 0,\; g_{\dot{r}}h + g h_{\dot{r}} = r',\; g_{r'}h + g h_{r'} = \dot{r},\\[0.15cm]
gg_r + hh_r  = -r(r'^2 + h^2),\;  gg_{\dot{r}} + r^2 h h_{\dot{r}} = -\dot{r},\\[0.15cm]
gg_{r'} + r^2 h h_{r'}  = -r^2r'
\end{array}
\end{equation}
which implies the `kinematical' relations
\begin{equation}\label{eq132} 
\begin{split}
\vec{m}_u & = \begin{pmatrix}
g_{\dot{r}} \\ h_{\dot{r}} 
\end{pmatrix} = \frac{1}{r^2h^2-g^2} 
\begin{pmatrix}
\dot{r} & r^2r' \\
 -r' & -\dot{r}
\end{pmatrix} 
\begin{pmatrix}
g \\ h
\end{pmatrix} =: K\vec{m} \\[0.15cm]
\vec{m}_w & = \begin{pmatrix}
g_{r'} \\ h_{r'} 
\end{pmatrix} = \frac{1}{r^2h^2-g^2} 
\begin{pmatrix}
r^2r' & r^2\dot{r} \\
 -\dot{r} & -r^2r'
\end{pmatrix} 
\begin{pmatrix}
g \\ h
\end{pmatrix} =: N\vec{m}\\[0.15cm]
\vec{m}_r & = \begin{pmatrix}
g_r \\ h_r
\end{pmatrix} = \frac{r^2(r'^2+h^2)}{r^2h^2-g^2} 
\begin{pmatrix}
1 & 0 \\
 0 & -1
\end{pmatrix} 
\begin{pmatrix}
g \\ h
\end{pmatrix} =: S\vec{m}
\end{split}
\end{equation}
$(r^2h^2-g^2)\cdot$\eqref{eq129} then becomes
\begin{equation}\label{eq133} 
(gr' + h\dot{r})(\ddot{r}-r^2r'') = -r(r'^2+h^2)(\dot{r}h - r'g).
\end{equation}
While for a moment one may wonder about the signs in front of $\dot{r}h$ and $gr'$ (as it would be so easy to simply divide by their sum resp. difference, if the relative signs on the 2 sides were the same), one again has to argue `with hindsight' in order to derive the desired conclusion 
\begin{equation}\label{eq134} 
D[r] := \ddot{r} - r^2 r'' - rr'^2 + rh^2 = 0,
\end{equation}
by writing \eqref{eq133} as
\begin{equation}\label{eq135} 
(gr' + h\dot{r})(D[r] + rr'^2 -rh^2) + r(r'^2 + h^2)(\dot{r}h - r'g) = 0
\end{equation}
and then showing that all terms {\it not} containing $D$ identically cancel (as well as noting that, as assumed, $gr' + h\dot{r} = \dot{z}r' - z'\dot{r} \neq 0$).\\
%%%%%%%%%%%%%%%%%%%%%%%%%%%%%%%%%%%%%%%%%%%%%%%%%%%%%%%%%%%%%%%%%%%%%%%%%%%%%
%PAGE 36 PAGE 36 PAGE 36 PAGE 36 PAGE 36 PAGE 36 PAGE 36 PAGE 36 PAGE 36 PAGE 36 
%%%%%%%%%%%%%%%%%%%%%%%%%%%%%%%%%%%%%%%%%%%%%%%%%%%%%%%%%%%%%%%%%%%%%%%%%%%%%
Analogously for
\begin{equation}\label{eq136} 
\tilde{g}' + \dot{\tilde{h}} = 0;
\end{equation}
assuming 
\begin{equation}\label{eq137} 
\begin{split}
\tilde{h}_w & = \frac{\partial \tilde{h}}{\partial r'} = -\frac{\partial \tilde{g}}{\partial \dot{r}} = - \tilde{g}_u, \\
\tilde{g}_w & = \frac{\partial \tilde{g}}{\partial r'} = -r^2\frac{\partial \tilde{h}}{\partial \dot{r}} = - r^2 \tilde{h}_u
\end{split}  ,
\end{equation}
which $\tilde{h} = f_{\pm}$ $(=\dot{z}_{\pm}=v'_{\pm}(t,\varphi))$ and  $\tilde{g} = r^2h = -r^2 z'_{\pm} = -\dot{v}_{\pm}$ do, \eqref{eq136} becomes (cp.\eqref{eq129})
\begin{equation}\label{eq138} 
\tilde{h}_{\dot{r}}(\ddot{r}-r^2r'') + \tilde{h}_r \dot{r} + \tilde{g}_r r' = 0.
\end{equation}
Now, however, assuming (cp. \eqref{eq130})
%%%%%%%%%%%%%%%%%%%%%%%%%%%%%%%%%%%%%%%%%%%%%%%%%%%%%%%%%%%%%%%%%%%%%%%%%%%%%
%PAGE 41' PAGE 41' PAGE 41' PAGE 41' PAGE 41' PAGE 41' PAGE 41' PAGE 41' PAGE 41' 
%%%%%%%%%%%%%%%%%%%%%%%%%%%%%%%%%%%%%%%%%%%%%%%%%%%%%%%%%%%%%%%%%%%%%%%%%%%%%
\begin{equation}\label{eq139} 
\begin{split}
\tilde{g}^2 + r^2\tilde{h}^2 & = r^2(1-r'^2) -r^2\dot{r}^2 = r^2[1-\dot{r}^2-r^2r'^2] \\
\tilde{g}\tilde{h} & = r^2(\dot{r}r')
\end{split}
\end{equation}
one gets
\begin{equation}\label{eq140} 
\begin{split}
\tilde{g}_r \tilde{h} + \tilde{g}\tilde{h}_r & = 2r\dot{r}r', \; \tilde{g}_{\dot{r}} \tilde{h} + \tilde{g}\tilde{h}_{\dot{r}} = r^2r', \; \tilde{g}_{r'} \tilde{h} + \tilde{g}\tilde{h}_{r'} = r^2\dot{r} \\
\tilde{g}\tilde{g}_r  + r^2\tilde{h}\tilde{h}_r & = r(1-\dot{r}^2-2r^2r'^2 -\tilde{h}^2),\; \tilde{g}\tilde{g}_{\dot{r}}  + r^2\tilde{h}\tilde{h}_{\dot{r}} = -\dot{r}r^2,\\
\tilde{g}\tilde{g}_{r'}  + r^2\tilde{h}\tilde{h}_{r'} & = -r^4r',
\end{split}
\end{equation}
which implies the (again, `kinematical') relations
\begin{equation}\label{eq141} 
\begin{pmatrix}
\tilde{g}_{\dot{r}} & \tilde{g}_{r'} & \tilde{g}_r \\
\tilde{h}_{\dot{r}} & \tilde{h}_{r'} & \tilde{h}_r
\end{pmatrix} = 
\frac{1}{r^2\tilde{h}^2 - \tilde{g}^2}
\begin{pmatrix}
r^2\tilde{h} & -\tilde{g} \\ -\tilde{g} & \tilde{h}
\end{pmatrix}
\begin{pmatrix}
r^2r' & r^2\dot{r} & 2r\dot{r}r' \\
-r^2\dot{r} & -r^4r' & r(1-\dot{r}^2 - 2r^2r'^2 - \tilde{h}^2)
\end{pmatrix}
\end{equation}
i.e.
\begin{equation}\label{eq142} 
\begin{split}
\vec{n}_u & = \begin{pmatrix}
\tilde{g}_{\dot{r}} \\ \tilde{h}_{\dot{r}} 
\end{pmatrix} = \frac{1}{r^2 \tilde{h}^2-\tilde{g}^2} 
\begin{pmatrix}
r^2\dot{r} & r^4r' \\
 -r^2r' & -r^2\dot{r}
\end{pmatrix} 
\begin{pmatrix}
\tilde{g} \\ \tilde{h}
\end{pmatrix} =: \tilde{K}\vec{n} \\[0.15cm]
\vec{n}_w & = \begin{pmatrix}
\tilde{g}_{r'} \\ \tilde{h}_{r'} 
\end{pmatrix} = \frac{1}{r^2 \tilde{h}^2-\tilde{g}^2} 
\begin{pmatrix}
r^4r' & r^4\dot{r} \\
 -r^2\dot{r} & -r^4r'
\end{pmatrix} 
\begin{pmatrix}
\tilde{g} \\ \tilde{h}
\end{pmatrix} =: \tilde{N}\vec{n} \\[0.15cm]
\vec{n}_r & = \begin{pmatrix}
\tilde{g}_{r} \\ \tilde{h}_{r} 
\end{pmatrix} = \frac{1}{r^2 \tilde{h}^2-\tilde{g}^2} 
\left( \begin{smallmatrix}
-r(1-\dot{r}^2-2r^2r'^2-\tilde{h}^2) & 2r^3\dot{r}r' \\
 -2r\dot{r}r' & {\small r(1-\dot{r}^2-2r^2r'^2-\tilde{h}^2)}
\end{smallmatrix} \right) 
\begin{pmatrix}
\tilde{g} \\ \tilde{h}
\end{pmatrix} =: \tilde{S}\vec{n}
\end{split}
\end{equation}
so that, in analogy with \eqref{eq133},
%%%%%%%%%%%%%%%%%%%%%%%%%%%%%%%%%%%%%%%%%%%%%%%%%%%%%%%%%%%%%%%%%%%%%%%%%%%%%
%PAGE 42 PAGE 42 PAGE 42 PAGE 42 PAGE 42 PAGE 42 PAGE 42 PAGE 42 PAGE 42 PAGE 42 
%%%%%%%%%%%%%%%%%%%%%%%%%%%%%%%%%%%%%%%%%%%%%%%%%%%%%%%%%%%%%%%%%%%%%%%%%%%%%
\begin{equation}\label{eq143} 
\begin{split}
r^2 (r'\tilde{g} + \dot{r}\tilde{h})(\ddot{r}-r^2r'') & = \dot{r}r\big[ -2\dot{r}\tilde{g}r' + (1-\dot{r}^2-2r^2r'^2-\tilde{h}^2)\tilde{h} \big]\\
& \quad + r'r \big[ -(1-\dot{r}^2 -2r^2r'^2-\tilde{h}^2)\tilde{g} + 2r^2\dot{r}r'\tilde{h} \big]
\end{split}
\end{equation}
is obtained; writing $\ddot{r}-r^2r'' = D[r] + rr'^2 - \frac{1}{r^3}\tilde{g}^2$, $r^2(r'\tilde{g}+\dot{r}\tilde{h})D[r] \stackrel{!}{=} 0$ is derived, as all terms not containing $D$ again identically cancel (note that the condition $r^2\tilde{h}^2-\tilde{g}^2 = -r^2(r^2h^2-g^2) \neq 0$, needed to obtain \eqref{eq132}/\eqref{eq142} from \eqref{eq131}/\eqref{eq140} was implicitly assumed from the start, as $\dot{z}^2 = r^2z'^2$ implies that one of $\sqrt{\pm} := \sqrt{1-(\dot{r}\pm rr')^2}$ vanishes, making derivatives of $f$, resp. $gh\tilde{g}\tilde{h}$ diverge there; apart from those critical points, however, \eqref{eq127} and \eqref{eq136} present 2, genuinely different, ways to obtain the (same) equation of motion, $D[r] = 0$).\\\\
In forthcoming work the progress made in understanding the dynamics of axially symmetric membranes will be applied to
general minimal hypersurfaces, Lorentzian and Euclidean.

\end{document}